\documentclass[preprint,a4paper]{elsarticle}

\usepackage[english]{babel}

\usepackage{geometry}

\usepackage{amssymb}
\usepackage{amsfonts}
\usepackage{amsmath}
\usepackage{latexsym}
\usepackage{mathtools}
\usepackage{bbm}

\usepackage{amsthm}

\usepackage{thmtools}
\usepackage{thm-restate}
\usepackage{environ}

\usepackage{url}
\usepackage{hyperref}
\usepackage{cleveref}

\usepackage{enumitem}

\usepackage{tikz}

\makeatletter
\newcommand*\bigcdot{\mathpalette\bigcdot@{.75}}
\newcommand*\bigcdot@[2]{\mathbin{\vcenter{\hbox{\scalebox{#2}{$\m@th#1\bullet$}}}}}
\makeatother

\newtheoremstyle{TheoremNum}
		{\topsep}{\topsep}					
		{\itshape}							
		{}								
		{\bfseries}							
		{.}								
		{ }								
		{\thmname{#1}\thmnote{ \bfseries #3}}	

\theoremstyle{TheoremNum}

\theoremstyle{plain}
\newtheorem{lemma}{Lemma}[section]

\newtheorem{corollary}[lemma]{Corollary}
\newtheorem{claim}[lemma]{Claim}

\theoremstyle{remark}

\newtheorem{remark}[lemma]{Remark}

\theoremstyle{definition}
\newtheorem{definition}[lemma]{Definition}
\newtheorem{example}[lemma]{Example}

\newenvironment{claimproof}{\begin{proof}\renewcommand{\qedsymbol}{\claimqed}}{\end{proof}\renewcommand{\qedsymbol}{\plainqed}}
\let\plainqed\qedsymbol

\DeclarePairedDelimiter\ceil{\lceil}{\rceil}
\DeclarePairedDelimiter\floor{\lfloor}{\rfloor}

\newcommand{\quantifierFormula}[3]{#1 #2 ,\, #3}

\newcommand{\forallFormula}[2]{\quantifierFormula{\forall}{#1}{#2}}

\newcommand{\existsFormula}[2]{\quantifierFormula{\exists}{#1}{#2}}

\def\BoolSet{\textsc{Bool}}
\def\True{\textsc{True}}
\def\False{\textsc{False}}

\def\ColorSet{\textsc{Colors}}

\def\WeightSet{\textsc{Weights}}
\def\wlesseq{\preceq}
\def\wless{\prec}
\def\wgteq{\succeq}
\def\minWeights{\min}
\def\Error{\textsc{Error}}
\def\weightsSum{\oplus}
\def\bigweightsSum{\bigoplus}

\def\sizeConstr{K}
\def\sizeConstrSet{\mathcal{\sizeConstr}}

\def\connectedComp{A}

\def\Conn{\textsc{Conn}}

\newcommand{\TvalidY}[1]{$T_{#1}$-valid}
\newcommand{\TvalidR}[1]{$T_{#1}$-partially valid}

\def\ColorPartitionSet{S}
\def\ColorAoneSet{F_1}
\def\ColorAfiveSet{F_5}

\newcommand{\F}{\mathbb{F}}
\newcommand{\N}{\mathbb{N}}

\newcommand{\X}{\mathcal{X}}
\newcommand{\C}{\mathcal{C}}

\newcommand{\T}{\mathcal{T}}
\newcommand{\B}{\mathcal{B}}
\newcommand{\R}{\mathcal{R}}

\newcommand{\set}[1]{\ensuremath{ \left\lbrace  #1 \right\rbrace }}

\newcommand{\partt}{\textrm{part}}

\newcommand{\snec}[1]{\textrm{snec}_{#1}}

\allowdisplaybreaks

\newcommand{\highlightChange}[1]{#1}

\begin{document}

\def\mainResultType{theorem}

\def\mainResultStatement{
\label{thm:mainResult}
Let $d$ and $q$ be positive integers.
For a $d$-stable locally checkable problem with $q$ colors, size constraints $\sizeConstrSet$, positive connectivity constraints $\C^+$ and negative connectivity constraints $\C^-$, there exists an algorithm that solves the problem on graphs with a given binary decomposition tree in time
$n^{O(w d q (\vert\C^+\vert + 1) 2^{\vert\C^-\vert})}$,
where $n$ is the number of vertices of the input graph and $w$ is the mim-width of the associated decomposition.
}

\begin{frontmatter}
\title{On \texorpdfstring{$d$}{d}-stable locally checkable problems parameterized by mim-width}

\def\BuenosAires{Buenos Aires, Argentina}
\def\UBA{Universidad de Buenos Aires}
\def\ICC{CONICET-\UBA. Instituto de Investigaci\'on en Ciencias de la Computaci\'on (ICC). \BuenosAires.}

\author[ICC]{Carolina Luc{\'i}a Gonzalez}
\ead{cgonzalez@dc.uba.ar}

\author[UNIFR]{Felix Mann\texorpdfstring{\corref{mycorrespondingauthor}}{}}
\cortext[mycorrespondingauthor]{Corresponding author}
\ead{felix.mann@unifr.ch}

\address[ICC]{\ICC}
\address[UNIFR]{University of Fribourg, Department of Informatics. Fribourg, Switzerland.}

\begin{abstract}
In this paper we continue the study of locally checkable problems under the framework introduced by Bonomo-Braberman and Gonzalez in 2020, by focusing on graphs of bounded mim-width.
We study which restrictions on a locally checkable problem are necessary in order to be able to solve it efficiently on graphs of bounded mim-width.
To this end, we introduce the concept of $d$-stability of a check function.
The related locally checkable problems contain large classes of problems, among which we can mention, for example, LCVP problems.
We give an algorithm showing that these problems
are XP when parameterized by the mim-width of a given binary decomposition tree of the input graph
\highlightChange{, that is, that they can be solved in polynomial time given a binary decomposition tree of bounded mim-width.}
We explore the relation between $d$-stable locally checkable problems and the recently introduced DN~logic (Bergougnoux, Dreier and Jaffke, 2022), and show that both frameworks model the same family of problems.
We include a list of concrete examples of $d$-stable locally checkable problems whose complexity on graphs of bounded mim-width was open so far.
\end{abstract}

\begin{keyword}
Locally checkable problem \sep
Mim-width \sep
d-stability \sep
Vertex partitioning problem \sep
DN logic \sep
Coloring \sep
Conflict-free coloring \sep
[k]-Roman domination \sep
b-coloring

\MSC[2010]
05C69 \sep
05C85 \sep
68Q25 \sep
68R10
\end{keyword}
\end{frontmatter}

\section{Introduction}
\label{sec:Introduction}
There are many graph problems which can be expressed as vertex coloring problems or, equivalently, as vertex partition problems.
Besides the well-known \textsc{Chromatic Number} and its variations, subset problems can be understood as a coloring problem with two colors, and path problems such as \textsc{Hamiltonian cycle} can be stated by asking for a vertex coloring with $n$ colors $1,\ldots,n$ (where $n$ is the number of vertices of some input graph) such that every vertex with color $i$ has to be adjacent to a vertex with color ${i-1}$ and a vertex of color ${i+1}$.
This broad variety of coloring problems makes it tempting not to try to solve each of these problems on its own but to find efficient algorithms for many of them at once.
The probably best-known meta-theorem in graph theory is Courcelle's theorem~\cite{COURCELLE199012}, which states that every graph problem expressible in Monadic Second-Order graph logic (MSO$_2$) can be solved in linear time on graphs of bounded treewidth.

\emph{Locally checkable problems} are a broad subclass of coloring problems.
A locally checkable problem is associated with a \emph{check function}, which is a function that takes as input a vertex $v$ of a graph $G$ and a coloring of the closed neighborhood of $v$ and outputs a Boolean value.
We are looking for a coloring $c$ of the vertices of $G$ such that the check function outputs \True\ for every vertex $v$ and the restriction of $c$ to $N[v]$.
Such a coloring will be called a \emph{proper coloring}, since this is a generalization of the notion of proper colorings in the problem \textsc{Chromatic Number}.
Indeed, determining the chromatic number can be seen as a locally checkable problem, where the check function simply verifies that the color of a vertex $v$ is unique in its closed neighborhood.
Another example for a locally checkable problem is the domination problem. 
Here, we are coloring the graph with two colors, $S$ and $\overline{S}$, and the check function verifies if the closed neighborhood of a vertex $v$ contains at least one vertex of color $S$.
Then, for any proper coloring, the vertices which have color $S$ form a dominating set.
Once a graph and a check function are given,
\highlightChange{we could ask not only whether there exists a proper coloring but also}
for an optimal proper coloring (for some notion of optimality).

Most locally checkable problems are NP-complete on general graphs.
For a given family of graphs it is thus interesting to determine under which conditions on the check function and the color set we can efficiently solve the associated locally checkable problems on that family.
In~\cite{LCPTreewidth} \highlightChange{Bonomo-Braberman and Gonzalez showed} that, under mild conditions on the check function\footnote{\highlightChange{Namely, the existence of \emph{polynomial partial neighborhood systems}. We omit the definition due to its technicality, but interested readers are encouraged to refer to~\cite{LCPTreewidth}.}}, locally checkable problems can be solved in polynomial time in graphs of bounded treewidth\footnote{\highlightChange{Some of these problems are in fact FPT parameterized by treewidth, while the others are simply XP.}}.
This line of research has been continued in~\cite{LCPcliquewidth} in which
\highlightChange{Baghirova, Gonzalez, Ries and Schindl}
specified a class of check functions called \emph{color-counting} check functions.
These functions only depend on the vertex, the color it receives and, for each color $a$, the number of neighbors of $v$ of color $a$. 
In other words, for a graph $G$ and a set of $q$ colors,
the check function can be written as a function $check$ which takes as input a vertex $v$, a color and, for each color $a$, a non-negative integer $k_{a}$.
In order to determine if a coloring $c$ is proper, we verify if $check(v,c(v),k_1,\ldots,k_q)=\True$ for every $v\in V(G)$, where $k_a = \left\vert\set{w \in V(G) : c(w) = a \text{ and $w$ is a neighbor of $v$}}\right\vert$.
\highlightChange{It is shown in~\cite{LCPcliquewidth} that} the problems associated to a color-counting check function can be solved in polynomial time on graphs of bounded clique-width if the number of colors $q$ is a constant.
If we ease these restriction\highlightChange{s} on the check function and on $q$, we obtain locally checkable problems which are NP-hard even on complete graphs (see~\cite{LCPTreewidth}) and thus, in this case, we do not expect to find a polynomial-time algorithm.

In this paper we turn our attention to graphs \highlightChange{which are bounded in another width parameter, the \emph{maximum induced matching width} (or \emph{mim-width} for short), which was introduced by Vatshelle in~\cite{VatshelleThesis} via the use of decomposition trees.
A \emph{binary decomposition tree} $(T,\delta)$ of $G$ consists of a rooted binary tree $T$ and a bijection $\delta$ between the leaves of $T$ and the vertices of $G$ (see~\cite[Definition 3.1.3]{VatshelleThesis}).
For any node $v$ of $T$ there is an associated partition of the vertices into two sets: the vertices corresponding to leaves which are descendants of $v$, denoted $T_v$, and all other vertices, denoted $T_{\overline{v}}$.
The \emph{mim-width of a graph} is the smallest number $k$ such that there is a binary decomposition tree $(T,\delta)$ of $G$ such that, for every node $v$ of $T$, the maximum induced matching in the bipartite graph $G[T_v,T_{\overline{v}}]$ has size at most $k$.
A more detailed definition is given in \Cref{sec:Preliminaries}.

In a sense, mim-width is a broader width parameter than treewidth or clique-width.
Indeed, any graph class of bounded clique-width has bounded mim-width, but there are graph classes of bounded mim-width whose clique-width is unbounded, such as interval graphs and permutation graphs~(\cite{GraphClassesWithStructuredNeighborhoods, cliquewidthInterval}).}
We observe that there are color-counting check functions with a bounded number of colors whose corresponding locally checkable problems are NP-hard on graphs of bounded mim-width (see \Cref{sec:Limits}).
Thus, we study which further restrictions we have to impose on a check function such that the associated locally checkable problems can be solved in polynomial time on graphs of bounded mim-width.

\highlightChange{Despite the broadness of families of bounded mim-width, there is still a surprising number of problems that can be solved in XP time when parameterized by mim-width.
In one of the first works to identify a large family of such problems (\cite{LCVSVP-paper2}),}
Bui-Xuan, Telle and Vatshelle focused on the following problems, defined \highlightChange{by Telle} in~\cite{LCVSVP-thesis}, which are related to a special type of color-counting check functions.
Given a positive integer $q$, a \emph{degree constraint matrix} $D$ is a $q \times q$-matrix whose entries are sets of non-negative integers.
A \emph{locally checkable vertex partitioning (LCVP)} problem is associated to a degree constraint matrix $D$ and consists in deciding whether there exists a partition $\{V_1, V_2, \dots, V_q\}$ of $V(G)$ such that for every $i,j\in[q]$ and every $v\in V_i$ it holds that $\vert N(v) \cap V_j\vert \in D[i,j]$.
\highlightChange{In~\cite{LCVSVP-thesis} there is also a broad list of problems given which can be expressed as LCVP problems, amongst which are Dominating set and Independent set.}
\highlightChange{In~\cite{gen-dist-dom-mim-width-2019}, Jaffke, Kwon, Str{\o}mme and Telle} showed that LCVP problems can be solved in polynomial time on graphs of bounded mim-width if $q$ is a constant and every entry of the associated degree constraint matrix is either finite or co-finite (that is, the complement of a finite set).
\highlightChange{Observe that all algorithms whose runtime is polynomial when the mim-width is bounded only work under the assumption that a suitable binary decomposition tree is given with the graph. 
While for some graph classes there are efficient algorithms to obtain such a decomposition (see, for example, \cite{GraphClassesWithStructuredNeighborhoods}), it is not known whether it is possible to obtain a binary decomposition tree of some graph $G$ with the smallest possible mim-width in polynomial time.
Even if we only ask for a binary decomposition tree of mim-width $f(mimw(G))$ for \emph{any} function $f$, we do not know if this is possible.}

Every LCVP problem can be considered as a locally checkable problem with a color-counting check function.
Indeed, given a degree constraint matrix $D$, we can associate the following color-counting check function to its corresponding LCVP problem:
\[
    check(v, i, k_1, \ldots, k_q)
    =
    \left(\forallFormula{j \in [q]}{k_j \in D[i,j]}\right).
\]
Observe that there are color-counting check functions which do not correspond to any LCVP problem.
For example, the ${k}$-domination problem asks for a coloring of the vertices of the graph with the numbers from 0 to $k$ such that, for each vertex $v$, the sum of the colors assigned to the vertices in the closed neighborhood of $v$ is at least $k$, and where we want to minimize the sum of the colors of all the vertices in the graph.
Another example is the conflict-free $q$-coloring, which asks for a coloring of the vertices with $q$ colors such that in the neighborhood of each vertex there is a color which appears exactly once.

In order to generalize LCVP problems to include problems as the ones above, we introduce the notion of a $d$\emph{-stable} check function.
\begin{definition}
Let $\set{a_1, \ldots, a_q}$ be a set of colors and let $check$ be a color-counting check function.
\highlightChange{Let $d$ be a positive integer.}
We say that $check$ is a \emph{$d$-stable check function} if
for every $v\in V(G)$ and any coloring $c$ of $V(G)$ we have
\[check(v, c(v), k_1, \ldots, k_q) = check(v, c(v), \min(d, k_1), \ldots, \min(d, k_q))\]
where
$k_j = \left\vert\set{w \in V(G) : c(w) = a_j \text{ and $w$ is a neighbor of $v$}}\right\vert$
for every $j \in [q]$.
\end{definition}
In other words, if a check function is $d$-stable that means that the ``properness'' of a coloring does not depend on the exact number of neighbors with a certain color as long as it is ``large enough'' (that is, larger than the threshold $d$).
For example, for the domination problem, we only need to know if every vertex has zero or at least one neighbor in the dominating set.
However, the exact number of neighbors in the set is not important if it is larger than one.
Thus, this problem has a 1-stable check function.

Observe that in the result of~\cite{LCVSVP-paper2} the condition that every entry of a degree constraint matrix $D$ is either finite or co-finite guarantees that its associated check function is $d$-stable for some $d$ (the value of $d$ depends on the matrix $D$, see~\cite{LCVSVP-paper2}). 
Thus, the notion of $d$-stability generalizes the set of LCVP problems which can be efficiently solved by the methods in~\cite{LCVSVP-paper2}.

We show that the $d$-stability of a check function suffices to yield an algorithm which solves the associated locally checkable problems in polynomial time on graphs of bounded mim-width (given a suitable binary decomposition tree).
We also show that it is possible to solve optimization versions of these problems, by considering a function which assigns weights depending on the coloring of the graph.
Given a coloring $c$ of $V(G)$ with colors in $\set{a_1, \ldots, a_q}$, a \emph{local weight function} assigns weights to vertices depending on the vertex and the coloring of its closed neighborhood.
Then the weight of the coloring $c$ is defined as the sum of the weights of all the vertices.
We consider local weight functions which have properties analogous to the ones previously defined for the check functions.
A \emph{color-counting weight function} is a local weight function where the assigned weights depend on the vertex $v$, the color it receives and, for each color $a$, the number of neighbors of $v$ which have color $a$.
\highlightChange{For a positive integer $d$,} we say that a color-counting weight function $\textsc{w}$ is $d$-stable if for every $v\in V(G)$ and every coloring $c$ we have $\textsc{w}(v, c(v), k_1, \ldots, k_q) = \textsc{w}(v, c(v),\min(d,k_1), \ldots, \min(d,k_q)),$
where $k_j = \vert N(v)\cap c^{-1}(a_j)\vert$ for every $j\in [q]$.
\highlightChange{A locally checkable problem with both a $d$-stable check function and a $d$-stable weight function will be called a \emph{$d$-stable locally checkable problem}.
We adapt the algorithm given in~\cite{LCVSVP-paper2} to show that finding an optimal proper coloring for such a problem is possible in polynomial time in graphs of bounded mim-width.
We also give the option to add global constraints on the sizes of the color classes and also connectivity constraints (among which we distinguish between \emph{positive} and \emph{negative} constraints, depending on whether they ask for a set of vertices to be connected or disconnected), for which we follow the ideas of Bergougnoux and Kanté in~\cite{bergougnoux2021applications}.

\begin{restatable}{\mainResultType}{mainResult}
\mainResultStatement
\end{restatable}}

We would like to analyze the relation of these $d$-stable functions with another subset of graph problems.
The \emph{distance neighborhood (DN) logic} was introduced by Bergougnoux, Dreier and Jaffke in~\cite{DN-logic} as an extension of existential MSO$_1$, that is, MSO$_1$ without universal quantifiers and without negation of terms containing existential quantifiers.
DN logic introduces further the use of \emph{neighborhood terms}, whose basic element is the following \highlightChange{\emph{neighborhood operator}}:
for a set $S$ of vertices of a graph, $N_d^r(S)$ denotes the set of every vertex $v$ for which there are at least $d$ vertices in $S$ whose distance to $v$ is at most $r$ and at least~1.
This logic allows us to combine these terms using the usual set operators as well as constant sets of vertices.
We can also use them in logical expressions by comparing sizes of these sets, asking for equality or if they are contained in each other.
A precise definition is given in~\cite{DN-logic}.
The A\&C DN logic is an extension of DN logic with the terms $\text{con(t)}$ and $\text{acy}(t)$, which state that the subgraph induced by the neighborhood term $t$ is connected or acyclic, respectively.
In \Cref{sec:LCPvsDN} we explore the modeling power of the frameworks.
For any \highlightChange{A\&C} DN expression
there is an equivalent
\highlightChange{formulation using $d$-stable locally checkable problems with size, connectivity and acyclicity constraints.}
\highlightChange{Conversely, we can model any $d$-stable locally checkable problem with such global constraints in A\&C DN logic.}
\highlightChange{It was also shown in~\cite{DN-logic}} that any A\&C DN formula can be solved in polynomial time on graphs of bounded mim-width, yielding \highlightChange{another polynomial} time algorithm solving $d$-stable locally checkable problems on graphs of bounded mim-width\highlightChange{, besides \Cref{thm:mainResult}}.
Since the two models look quite different, it may seem surprising that they can solve the same set of problems.
However, this astonishment is dampened given the fact that they are both generalizations of the proof idea in \cite{LCVSVP-paper2}.

\highlightChange{T}he additional translation step comes at the expense of a worse runtime.
For a $d$-stable coloring problem on $q$ colors without connectivity or acyclicity constraints and an $n$-vertex graph $G$ whose mim-width is $w$, the algorithm in \Cref{sec:Algorithm} has a runtime of $n^{O(dwq)}$.
\highlightChange{Using the proof in \Cref{sec:LCPtoDN} and \cite[Theorem 1.2]{DN-logic} gives an algorithm of complexity $n^{O(w q^2 d^2 (d+1)^{2q})}$.}
Clearly, the first algorithm has a better exponent.
In some cases, it is possible to find shorter DN-expressions for a locally checkable problem than the one\highlightChange{s} we obtain from \Cref{sec:LCPtoDN}. 
For example, in \cite[Section 6.3]{DN-logic},
\highlightChange{there is}
a DN-formula expressing the conflict-free $q$-coloring problem which has length $O(q^2)$.
This is shorter than $O(q^2 2^{2q+1})$, the length we obtain from \Cref{sec:LCPtoDN}, but this still yields an algorithm of runtime $n^{O(dwq^2)}$ and thus a slower one than \Cref{sec:Algorithm}.

Finally, it is certainly of interest to determine if the restrictions we put on the check and weight function as well as on the number of colors could be eased.
In \Cref{sec:Limits} we show that this is not possible in general.
In particular, we show that there are locally checkable problems with non $d$-stable color-counting check functions which are NP-hard even on interval graphs.
Thus, we cannot expect to remove the condition of $d$-stability.
We also give locally checkable problems with $d$-stable check function for which the associated optimization problem is NP-hard for certain non $d$-stable \highlightChange{color-counting} weight functions.

Our paper is structured as follows.
We begin by explaining basic notions and definitions in \Cref{sec:Preliminaries}.
The definition and notation of $d$-stable locally checkable problems is given in \Cref{sec:d-stableLCP}.
In \Cref{sec:LCPvsDN} we analyze the relation between DN logic and $d$-stable locally checkable problems.
We then give a proof for our main result, \Cref{thm:mainThm}, in \Cref{sec:Algorithm}.
\Cref{sec:Limits} is dedicated to showing that the restrictions we give on the number of colors and on the check and weight functions are in some sense tight.
Finally, \Cref{sec:Applications} provides examples of problems which are associated to a $d$-stable check function and which have not been solved before on graphs of bounded mim-width.

\section{Preliminaries}
\label{sec:Preliminaries}
\label{sec:prelimAlgebra}

Let $k$ be a
\highlightChange{positive}
integer
and denote by $[k]$ the set $\set{1, \ldots, k}$.
We denote by $\N$ the set of non-negative integers, and by $\F_2$ the field with two elements.
For a set $S$, $\mathcal{P}(S)$ denotes the set of all subsets of $S$.

Throughout this paper we will work with the set $\BoolSet = \{\True, \False\}$ of Boolean values (truth values), and with all the usual logical operators, such as $\neg$, $\land$, $\lor$ and $\Rightarrow$, as well as with the quantifiers $\forall, \exists$.
\highlightChange{We use the following notation style for quantifiers: $\forallFormula{x \in X}{P}$ and $\existsFormula{x \in X}{P}$.}
Let $\mathbbm{1} \colon \BoolSet \rightarrow \F_2$ be the function such that $\mathbbm{1}(\True) = 1$ and $\mathbbm{1}(\False) = 0$.

Let $k$ and $q$ be non-negative integers.
We denote by $\partt_q(k)$ the set of all $q$-tuples of non-negative integers such that the sum of all entries is $k$.
For a set $S$, we denote by $\partt_q(S)$ the set of ordered partitions of $S$ into $q$ parts.
For some $X \in \partt_q(S)$, we denote by $\vert X\vert$ the $q$-tuple with $\vert X\vert_i = \vert X_i\vert$.
If $X \in \partt_q(S)$ and $S'$ is another set then we denote by $S' \cap X$ the $q$-tuple with $(S' \cap X)_i = S' \cap X_i$.
If $S$ and $S'$ are disjoint sets and $X \in \partt_q(S)$, $X' \in \partt_q(S')$ then we denote by $X \cup X'$ the entry-wise union of $X$ and $X'$, which is a partition of $S\cup S'$.

When $S$ and $S'$ are sets, an \emph{$S$-tuple with entries from $S'$} is a map from $S$ to $S'$.
Given an $S$-tuple $T$ for which every $x\in S$ is mapped to $T_x$, we sometimes write $T = (T_x)_{x\in S}$.

\subsection{Basic graph theory definitions}
\label{sec:prelimGraphs}

Throughout the whole paper, we assume that all graphs are finite, simple and undirected, unless stated differently.

Given a graph $G$, we denote with $V(G)$ the vertex set of $G$ and with $E(G)$ its edge set.
For a vertex $v \in V(G)$, the \emph{open neighborhood of $v$} ($N_G(v)$) is the set of neighbors of $v$ in $G$, and the \emph{closed neighborhood of $v$} is $N_G[v] = N_G(v) \cup \{v\}$.
The \emph{closed neighborhood of a set} $S \subseteq V(G)$ is $N_G[S] = \bigcup_{v \in S} N_G[v]$.
The \emph{degree} of a vertex $v$ is $d_G(v) = \vert N_G(v)\vert$.
We may omit the sub-index $G$ when it is clear from the context.

Given two disjoint sets $S, S' \subseteq V(G)$, we denote by $G[S]$ the subgraph of $G$ induced by $S$, and by $G[S,S']$ the bipartite graph of vertex set $S \cup S'$ and edge set $\set{\highlightChange{vv'} \in E(G) : v \in S, v' \in S'}$.

\highlightChange{A graph $G$ is \emph{connected} if for every pair of vertices $u,v \in V(G)$ there exists a path in $G$ from $u$ to $v$.
A \emph{connected component} of a graph is the set of vertices of an inclusion-wise maximal connected subgraph.
For two vertices $u,v$ in a connected graph $G$, the \emph{distance between $u$ and $v$} is the number of edges in a shortest path from $u$ to $v$ in $G$, and we denote it by $\text{dist}_{G}(u, v)$.}
Let $cc(G)$ denote the set of connected components of a graph $G$.
We define $ccut(G) = \partt_2(cc(G))$.
An element of $ccut(G)$ is called a \emph{connected cut}.
For a set $S \subseteq V(G)$, let $ccut(S) = ccut(G[S])$.

\highlightChange{Let $r$ be a positive integer.
The \emph{$r$-th power} of a graph $G$, denoted $G^r$, is the graph of vertex set $V(G)$ such that for all distinct vertices $u, v \in V(G)$, $u$ is adjacent to $v$ in $G^r$ if and only if $\text{dist}_{G}(u, v) \leq r$.
The \emph{$r$-neighborhood} $N^r_G(v)$ of a vertex $v$ in $G$ is the set of vertices different from $v$ which are at distance at most $r$ to $v$ in $G$.
Equivalently, this is the neighborhood of $v$ in the $r$-th graph power of $G$.
That is, $N_{G}^{r}(v) = N_{G^r}(v)$.}

A \emph{rooted tree} is a tree with a distinguished vertex called \emph{root}.
We will refer to the vertices of rooted trees as \emph{nodes}.
A \emph{rooted binary tree} is a rooted tree in which every non-leaf node has two children.
For a graph $G$, a rooted binary tree $T$ together with a bijection $\delta$ between $V(G)$ and the leafs of $T$ is called a \emph{binary decomposition tree} of $G$~\highlightChange{(\cite[Definition 3.1.3]{VatshelleThesis})}.
Let $(T, \delta)$ be a binary decomposition tree of a graph $G$.
Let $v \in V(T)$, we denote by $T_v$ the set of vertices of $G$ which correspond to the leaves which are descendants of $v$.
We denote $T_{\overline{v}} = V(G) \setminus T_v$.

Let $mim(G,S)$ denote the maximum size of an induced matching in the bipartite graph $G[S, V(G) \setminus S]$ (measured as number of edges), and let $mimw(T, \delta) = \max_{v \in V(T)}\set{mim(G,T_v)}$.
The mim-width of a graph $G$, denoted $mimw(G)$, is the minimum value of $mimw(T, \delta)$ over all binary decomposition trees $(T, \delta)$ of $G$.
Given a graph class $\mathcal{G}$, we say that $\mathcal{G}$ is of \emph{bounded mim-width} if $\sup\set{mimw(G) : G \in \mathcal{G}} < \infty$.
All graphs classes of bounded clique-width have bounded mim-width, but there exist graphs classes (for example, interval graphs) of mim-width 1 and unbounded clique-width~\cite{VatshelleThesis,cliquewidthInterval}.

\subsection{Weight sets}
\label{sec:prelimWeights}

Let $(\WeightSet, \wlesseq)$ be a totally ordered set with a maximum element (called $\Error$), together with the minimum operation of the order $\wlesseq$ (called $\minWeights$) and a closed binary operation on $\WeightSet$ (called $\weightsSum$) that is commutative and associative, has a neutral element and an absorbing element that is equal to $\Error$, and the following property is satisfied: $s_1 \wlesseq s_2 \Rightarrow s_1 \weightsSum s_3 \wlesseq s_2 \weightsSum s_3$ for all $s_1, s_2, s_3 \in \WeightSet$.
In such a case, we say that $(\WeightSet, \wlesseq, \weightsSum)$ is a \emph{weight set}.

A classic example of a weight set is $(\mathbb{N} \cup \set{+\infty}, \leq, +)$.
Notice that the maximum element is $+\infty$ in this case.
We could also consider the reversed order of natural weights: $(\mathbb{N} \cup \set{-\infty}, \geq, +)$, where the maximum element is now $-\infty$.
Another simple example worth mentioning is $(\set{0,1}, \leq, \max)$.

\section{\texorpdfstring{$d$}{d}-stable locally checkable problems}
\label{sec:d-stableLCP}
In this paper we will consider the following type of problems on graphs, which we will call \emph{$d$-stable locally checkable problems}.
Let $d$ and $q$ be positive integers and suppose we are given:
\begin{itemize}
\item a simple undirected graph $G$,

\item a set $\ColorSet = \{a_1, \ldots, a_q\}$ of $q$ colors,

\item a weight set $(\WeightSet, \wlesseq, \weightsSum)$,

\item a $d$-stable weight function $\textsc{w} \colon V(G) \times \ColorSet \times \N^{q} \to \WeightSet$, and

\item a $d$-stable check function $check \colon V(G) \times \ColorSet \times \N^{q} \to \BoolSet$.
\end{itemize}

We say that a coloring $c \colon V(G) \to \ColorSet$ of the input graph $G$ is a \emph{proper coloring} if for every vertex $v \in V(G)$ we have that $check(v, c(v), k_1, \ldots, k_q) = \True$, where $k_j = \vert\set{u \in N_G(v) : c(u) = a_j}\vert$ for all $j \in [q]$.
The \emph{weight of a coloring} $c$ is defined as $\textsc{w}(c) = \bigweightsSum_{v \in V(G)} \textsc{w}(v, c(v), k_1, \ldots, k_q)$ where $k_j = \vert\set{u \in N_G(v) : c(u) = a_j}\vert$ for all $j \in [q]$.
The goal is to find the minimum\footnote{According to the order $\wlesseq$ of the weight set.} weight of a proper coloring of the input graph.
Notice that if no such coloring exists then the answer is the maximum element of the weight set.

\highlightChange{
Observe that a $d$-stable check function has $\vert V(G)\vert q (d+1)^q$ different inputs, so its runtime could be dependent on $\vert V(G)\vert$.
However, throughout this paper we implicitly demand that the runtime of every check function only depends on $d$ and $q$.
Furthermore, we require that applying $\weightsSum$ is a constant time operation.
}

We can further consider size, connectivity and acyclicity constraints:
\begin{itemize}
\item a set $\sizeConstrSet \subseteq \partt_q(\vert V(G)\vert)$ of size constraints,

\item \highlightChange{a set $\mathcal{C}^{+} \subseteq \mathcal{P}(\ColorSet)$ of positive connectivity constraints and a set $\mathcal{C}^{-} \subseteq \mathcal{P}(\ColorSet)$ of negative connectivity constraints,}

\item \highlightChange{a set $\mathcal{A}^{+} \subseteq \mathcal{P}(\ColorSet)$ of positive acyclicity constraints and a set $\mathcal{A}^{-} \subseteq \mathcal{P}(\ColorSet)$ of negative acyclicity constraints,}
\end{itemize}
and we ask for the minimum weight of a proper coloring $c$ of the vertices of $G$ such that:
\highlightChange{\begin{itemize}
\item {\color{black}the $q$-tuple $(\vert\set{v \in V(G) : c(v) = a_1}\vert, \ldots, \vert\set{v \in V(G) : c(v) = a_q}\vert)$ is in $\sizeConstrSet$ (that is, the sizes of the $q$ color classes are ``acceptable''),}
\item the subgraph of $G$ induced by the set $\set{v \in V(G) : c(v) \in C}$ is connected (respectively not connected) for every $C \in \mathcal{C}^{+}$ (respectively $\mathcal{C}^{-}$),
and also
\item the subgraph of $G$ induced by the set $\set{v \in V(G) : c(v) \in A}$ is acyclic (respectively not acyclic) for every $A \in \mathcal{A}^{+}$ (respectively $\mathcal{A}^{-}$).
\end{itemize}}
Notice that $\mathcal{K}$ can capture \emph{any} conditions on the size\highlightChange{s} of the color classes.
We can therefore more naturally implicitly define the set $\mathcal{K}$ by stating these conditions in words.
The same holds for $\mathcal{C}$ and $\mathcal{A}$.

In order to illustrate \highlightChange{these definitions} better, we present the following examples of well-known problems modeled as $d$-stable locally checkable problems (with or without additional global constraints).

\begin{example}
Consider the {\sc Maximum Independent Set} problem.
This problem can be seen as a $1$-stable locally checkable problem with two colors:
\begin{itemize}
	\item $\ColorSet = \{\textsc{s}, \overline{\textsc{s}}\}$,

	\item ($\WeightSet, \wlesseq, \weightsSum) = (\mathbb{N} \cup \set{-\infty}, \geq, +)$,

	\item $\textsc{w}(v, \textsc{s}, k_{\textsc{s}}, k_{\overline{\textsc{s}}}) = 1$ and $\textsc{w}(v, \overline{\textsc{s}}, k_{\textsc{s}}, k_{\overline{\textsc{s}}}) = 0$ for all $v \in V(G)$ and $k_{\textsc{s}}, k_{\overline{\textsc{s}}} \in \N$,

	\item $check(v, a, k_{\textsc{s}}, k_{\overline{\textsc{s}}}) = (a = \overline{\textsc{s}} \lor k_{\textsc{s}} = 0)$ for all $v \in V(G)$, $a \in \ColorSet$ and $k_{\textsc{s}}, k_{\overline{\textsc{s}}} \in \N$.
\end{itemize}

In a proper coloring $c$, the set of vertices receiving color $\textsc{s}$ is independent by definition of the check function.
Conversely, every independent set defines a proper coloring.
In addition, by definition of the weights, $\textsc{w}(c) = \vert \set{v \in V(G) : c(v) = \textsc{s}}\vert$.
Therefore, the minimum weight of a proper coloring equals the maximum size of an independent set.
\end{example}

\begin{example}
The {\sc $q$-Equitable Coloring} problem can be modeled as a $1$-stable locally checkable problem with $q$ colors:
\begin{itemize}
	\item $\ColorSet = [q]$,

	\item ($\WeightSet, \wlesseq, \weightsSum) = (\set{0,1}, \leq, \max)$,

	\item $\textsc{w}(v, a, k_1, \ldots, k_q) = 0$ for all $v \in V(G), a \in \ColorSet$ and $k_1, \ldots, k_q \in \N$,

    \item $check(v, a, k_1, \ldots, k_q) = (k_a = 0)$ for all $v \in V(G)$, $a \in \ColorSet$ and $k_1, \ldots, k_q \in \N$,

	\item for all $j \in [q]$, the size of the color class $j$ is $\floor{\frac{n}{q}}$ or $\ceil{\frac{n}{q}}$.
\end{itemize}

In this case, we are not interested in the weight of the coloring but only in the existence of it.
Thus, the weights here serve the sole purpose of distinguishing if the solution is the maximum element $1$ (there does not exist any proper coloring satisfying the size constraints) or $0$ (such a coloring indeed exists).
And it is easy to see that a $q$-equitable coloring exists if and only if a proper coloring exists.
\end{example}

\begin{example}
The {\sc Minimum Connected Dominating Set} problem can also be expressed as a $1$-stable locally checkable problem with two colors:
\begin{itemize}
	\item $\ColorSet = \{\textsc{s}, \overline{\textsc{s}}\}$,

	\item ($\WeightSet, \wlesseq, \weightsSum) = (\mathbb{N} \cup \set{+\infty}, \leq, +)$,

	\item $\textsc{w}(v, \textsc{s}, k_{\textsc{s}}, k_{\overline{\textsc{s}}}) = 1$ and $\textsc{w}(v, \overline{\textsc{s}}, k_{\textsc{s}}, k_{\overline{\textsc{s}}}) = 0$ for all $v \in V(G)$ and $k_{\textsc{s}}, k_{\overline{\textsc{s}}} \in \N$,

	\item $check(v, a, k_{\textsc{s}}, k_{\overline{\textsc{s}}}) = (a = \textsc{s} \lor k_{\textsc{s}} = 1)$ for all $v \in V(G)$, $a \in \ColorSet$ and $k_{\textsc{s}}, k_{\overline{\textsc{s}}} \in \N$,

	\item the color class $\set{\textsc{s}}$ is connected.
\end{itemize}

In a proper coloring, the set of vertices receiving color $\textsc{s}$ is dominating by definition of $check$.
Furthermore, by considering $\C = \set{\set{\textsc{s}}}$, we are asking that the set of vertices with color $\textsc{s}$ induces a connected subgraph.
Conversely, every connected dominating set defines a proper coloring satisfying the connectivity constraint.
Since the weights of vertices with color $\textsc{s}$ is 1 and the weight of those with color $\overline{\textsc{s}}$ is 0, then the weight of the coloring is equivalent to the size of vertices receiving color $\textsc{s}$.
Therefore, the minimum weight of a proper coloring satisfying the connectivity constraints equals the minimum size of a connected dominating set.
\end{example}

More examples of $d$-stable locally checkable problems with size and connectivity constraints can be found in \Cref{sec:Applications}.
Many of the examples mentioned in~\cite[Table~1]{LCPTreewidth} are also $d$-stable (although this is not explicitly mentioned).
It should further be noted that the ``list versions'' of these problems, where each vertex $v$ has a list $L_v$ of possible colors that it can receive, can be modeled by adding the condition ``$a \in L_v$'' in the definition of $check(v, a, k_1, \ldots, k_q)$ for every vertex $v$.

\highlightChange{We can also consider problems where the local property involves not only the neighbors but vertices at larger distances.
Let $t$ be a fixed positive integer and let $r_1, \ldots, r_t$ be positive integers.
We can modify the domain of $d$-stable check and weight functions to be $V(G) \times \ColorSet \times \N^{qt}$ instead of $V(G) \times \ColorSet \times \N^{q}$, where $q$ is the number of colors.
That is, instead of having an input vector $k = (k_1, \ldots, k_q)$ where each $k_j$ indicates the number of neighbors of color $a_j$, we now have $\kappa = ((\kappa^1_1, \ldots, \kappa^1_q), \ldots, (\kappa^t_1, \ldots, \kappa^t_q))$ where each $\kappa^i_j$ indicates the number of vertices of color $a_j$ that are at distance at most $r_i$.
In this way, we have a check function $check \colon V(G) \times \ColorSet \times \N^{qt} \to \BoolSet$ and say that a coloring $c$ is a proper coloring if for every vertex $v \in V(G)$ we have that $check(v, c(v), \kappa) = \True$, where $\kappa^i_j = \vert\set{u \in N_{G}^{r_i}(v) : c(u) = a_j}\vert$ for all $i \in [t]$ and $j \in [q]$.
The weight function can be extended analogously.
We say that these functions are \emph{$d$-stable for $t$ distances $r_1, \ldots, r_t$}.
A locally checkable problem with such functions will be called \emph{$d$-stable locally checkable for $t$ distances $r_1, \ldots, r_t$}.
We may omit mentioning $r_1, \ldots, r_t$ when they are irrelevant.}

\section{Comparison with \highlightChange{A\&C} DN logic}
\label{sec:LCPvsDN}
In this section we show that \highlightChange{the set of problems that can be modeled using A\&C DN logic formulas of~constant length is equivalent to the set of problems that can be modeled using locally checkable problems with constant number of colors, functions that are $d$-stable for $t$ distances (for some constant positive integers $d$ and $t$), and size, connectivity and acyclicity constrains.}

\highlightChange{The \emph{distance neighborhood (DN) logic}, defined in~\cite{DN-logic}, is obtained by extending existential MSO$_1$ with a \emph{neighborhood operator $N_d^r(\cdot)$}:
for a term $t$ that represents a set $U$ of vertices in a graph $G$, the term $N_d^r(t)$ represents the set of all vertices in $G$ which are at distance at most $r$ and at least 1 from at least $d$ vertices in $U$.
The DN logic also includes size measurement of terms ($|\cdot| \leq \cdot$) and comparison between terms (with $=$ and $\subseteq$).
For a DN logic formula $\varphi$, $d(\varphi)$ denotes the greatest integer $d$ such that there is an operator $N_d^{\cdot}(\cdot)$ in $\varphi$.
Further, $R(\varphi)$ denotes the set of integers $r$ such that there is an operator $N_{\cdot}^{r}(\cdot)$ in $\varphi$.

The \emph{A{\&}C DN logic} is a further extension of DN logic with the formulas $\text{con}(\cdot)$ and $\text{acy}(\cdot)$.
Here, the expression $\text{con}(t)$ ($\text{acy}(t)$, respectively) represents that $G[U]$ is connected (acyclic, respectively), where $U$ is the set of vertices represented by $t$.

It was shown in~\cite{DN-logic} that any A\&C DN formula $\varphi$ can be expressed as an \emph{A\&C CDN formula} $\varphi'$, which is a formula of the form $\existsFormula{X_1 \ldots X_k}{\psi}$ where $\psi$ is a Boolean combination (that is, an expression using $\land$, $\lor$ and $\lnot$) of the following types of \emph{primitive formulas}:
\begin{enumerate}[label=\arabic*.]
    \item $(P = X_i)$, for a fixed subset $P$ and some variable $X_i$,

    \item $(X_i = X_j)$, for some variables $X_i, X_j$,
    
    \item $(X_i = \overline{X_j})$, for some variables $X_i, X_j$,
    
    \item $(X_i \cap X_j = X_h)$, for some variables $X_i, X_j, X_h$,
    
    \item $(N_d^r(X_i) = X_j)$, for some variables $X_i, X_j$ and a positive integer $d$,
    
    \item $(\vert X_i\vert \leq m)$, for some variable $X_i$ and a positive integer $m$,
    
    \item $\text{acy}(X_i)$, for some variable $X_i$,

    \item $\text{con}(X_i)$, for some variable $X_i$.
\end{enumerate}
Moreover, the length of $\varphi'$ is at most ten times the length of $\varphi$, $d(\varphi') = d(\varphi)$ and $R(\varphi')=R(\varphi)$~(\cite[Observation 3.1]{DN-logic}).}

\subsection{DN logic formulas as \texorpdfstring{$d$}{d}-stable locally checkable problems}
\label{sec:DNtoLCP}
In the following we will show how to model \highlightChange{an A\&C DN formula with $d$-stable locally checkable problems for $t$ distances, for some constants $t$ and $d$, with size, connectivity and acyclicity constraints.}
\highlightChange{\begin{lemma}
Let $\Pi$ be a problem that can be expressed by an A\&C DN formula $\varphi$.
Then, there exist some positive integers $p \leq 2^{10 \vert\varphi\vert}$ and $t \leq \vert\varphi\vert$, and a set $\{\Pi_1, \ldots, \Pi_{p}\}$ of $d(\varphi)$-stable locally checkable problems for $t$ distances with size, connectivity and acyclicity constraints, such that, for a given input, the optimal among all the optimal solutions of $\Pi_1, \ldots, \Pi_{p}$ is equivalent to the optimal solution of $\Pi$.
\end{lemma}}
\begin{proof}
\highlightChange{Let $G$ be the input graph and $\phi$ an A\&C CDN formula equivalent to $\varphi$, where $\phi = (\existsFormula{X_1 \ldots X_k}{\psi})$ and $\psi$ is a Boolean combination of primitive formulas of the eight types given above.}
Let $A = A_1 \cup \cdots \cup A_{\highlightChange{8}}$ be the set of all of such primitive formulas in $\psi$, where $A_i$ contains exactly the primitive formulas which are of type $i$, for every $i \in [\highlightChange{8}]$.

There are at most $\vert\psi\vert$ primitive formulas in $A$ and thus at most $2^{\vert\psi\vert}$ truth assignments $T \colon A \to \BoolSet$.
For every truth assignment that satisfies $\psi$, we will show how to model an equivalent check function and size restrictions of color classes.
Let \highlightChange{$T$} be a satisfying truth assignment.
For every $i \in [\highlightChange{8}]$, let $A_i^+$ be the set of all primitive formulas from $A_i$ which are set to \True\ by $\highlightChange{T}$.
Set $A_i^- = A_i \setminus A_i^+$.
In other words, the set $A_i^+$ contains all the primitive formulas of $A_i$ which need to hold in order to satisfy $\psi$ following the truth assignment $T$.
Similarly, $A_i^-$ contains all the formulas of $A_i$ that must not hold according to~$T$.

\highlightChange{Let $R$ be $R(\varphi)$ if this set is non-empty, otherwise we define $R$ as $\{1\}$.
Notice that the size of $R$ is no longer than the length of $\varphi$.}

Set $\ColorSet = \mathcal{P}([k]) \times \mathcal{P}(A_1) \times \mathcal{P}(A_5)$.
Intuitively, if a vertex $v$ obtains the color $(\ColorPartitionSet, \ColorAoneSet, \ColorAfiveSet)$, then
$\ColorPartitionSet$ represents the subset of the variables $X_1, \ldots, X_k$ in which $v$ is contained, and
$\ColorAoneSet$ and $\ColorAfiveSet$ save which of the formulas in $A_1$ and $A_5$, respectively, are satisfied locally at $v$.
We will define a check function that only makes sure that $\ColorAoneSet$ and $\ColorAfiveSet$ are set correctly according to this idea.
\highlightChange{This function will be $d(\varphi)$-stable for $|R|$ distances $r \in R$.}
For every $v \in V(G)$, $(\ColorPartitionSet, \ColorAoneSet, \ColorAfiveSet) \in \ColorSet$ and \highlightChange{$\ell \in \N^{|\ColorSet| \times |R|}$
(whose entries we denote by $\ell^{r}_{(\ColorPartitionSet', \ColorAoneSet', \ColorAfiveSet')}$ for any $r \in R$ and $(\ColorPartitionSet', \ColorAoneSet', \ColorAfiveSet') \in \ColorSet$),}
set $check(v, (\ColorPartitionSet, \ColorAoneSet, \ColorAfiveSet), \highlightChange{\ell}) = \True$ if and only if the following two statements hold:
\begin{itemize}
\item for all $(P = X_i) \in A_1$, we have that $(P = X_i)\in \ColorAoneSet$ if and only if $(i \in \ColorPartitionSet)$ and $(v \in P)$ are either both true or both false,
\item for every $(N_d^{\highlightChange{r}}(X_i) = X_j) \in A_5$, we have that $(N_d^{\highlightChange{r}}(X_i) = X_j) \in \ColorAfiveSet$ if and only if $(j \in \ColorPartitionSet)$ and $\left(\sum_{\ColorAoneSet' \subseteq A_1} \sum_{\ColorAfiveSet' \subseteq A_5} \sum_{\ColorPartitionSet_i \subseteq [k], i \in \ColorPartitionSet_i} \ell^{\highlightChange{r}}_{\ColorPartitionSet_i, \ColorAoneSet', \ColorAfiveSet'} \geq d\right)$ are both true or both false.
\end{itemize}

We define the following size\highlightChange{, connectivity and acyclicity} constraints which then guarantee that $\psi$ is satisfied:
\begin{itemize}
   \item For every $(P = X_i) \in A_1^+$
   and for every $(\ColorPartitionSet, \ColorAoneSet, \ColorAfiveSet) \in \ColorSet$ where $\ColorAoneSet$ does not contain $(P = X_i)$,
   the size of the color class $(\ColorPartitionSet, \ColorAoneSet, \ColorAfiveSet)$ is 0.

    \item For every $(P = X_i) \in A_1^-$,
    there exists a color $(\ColorPartitionSet, \ColorAoneSet, \ColorAfiveSet) \in \ColorSet$ such that $\ColorAoneSet$ does not contain $(P = X_i)$
    and the size of the color class $(\ColorPartitionSet, \ColorAoneSet, \ColorAfiveSet)$ is at least 1.

    \item For every $(X_i = X_j) \in A_2^+$
    and for every $(\ColorPartitionSet, \ColorAoneSet, \ColorAfiveSet) \in \ColorSet$ where $\vert \ColorPartitionSet \cap \{i,j\}\vert = 1$,
    the size of the color class $(\ColorPartitionSet, \ColorAoneSet, \ColorAfiveSet)$ is 0.

    \item For every $(X_i = X_j) \in A_2^-$,
    there exists a color $(\ColorPartitionSet, \ColorAoneSet, \ColorAfiveSet) \in \ColorSet$ such that $\vert \ColorPartitionSet \cap \{i,j\}\vert = 1$
    and the size of the color class $(\ColorPartitionSet, \ColorAoneSet, \ColorAfiveSet)$ is at least 1.

    \item For every $(X_i = \overline{X_j}) \in A_3^+$
    and for every $(\ColorPartitionSet, \ColorAoneSet, \ColorAfiveSet) \in \ColorSet$ where $\vert \ColorPartitionSet \cap \{i,j\}\vert \neq 1$,
    the size of the color class $(\ColorPartitionSet, \ColorAoneSet, \ColorAfiveSet)$ is 0.

    \item For every $(X_i = \overline{X_j}) \in A_3^-$,
    there exists a color $(\ColorPartitionSet, \ColorAoneSet, \ColorAfiveSet) \in \ColorSet$ such that $\vert \ColorPartitionSet \cap \{i,j\}\vert \neq 1$
    and the size of the color class $(\ColorPartitionSet, \ColorAoneSet, \ColorAfiveSet)$ is at least 1.

    \item For every $(X_i \cap X_j = X_h) \in A_4^+$
    and for every $(\ColorPartitionSet, \ColorAoneSet, \ColorAfiveSet) \in \ColorSet$ where
    $\ColorPartitionSet \cap \{i,j,h\}$ is either $\{i,j\}$, $\{i,h\}$, $\{j,h\}$, or $\{h\}$,
    the size of the color class $(\ColorPartitionSet, \ColorAoneSet, \ColorAfiveSet)$ is 0.

    \item For every $(X_i \cap X_j = X_h) \in A_4^-$,
    there exists a color $(\ColorPartitionSet, \ColorAoneSet, \ColorAfiveSet) \in \ColorSet$ such that
    $\ColorPartitionSet \cap \{i,j,h\} \in \{\{i,j\}, \{i,h\}, \{j,h\}, \{h\}\}$
    and the size of the color class $(\ColorPartitionSet, \ColorAoneSet, \ColorAfiveSet)$ is at least 1.

    \item For every $(N_d^{\highlightChange{r}}(X_i) = X_j) \in A_5^+$
    and for every $(\ColorPartitionSet, \ColorAoneSet, \ColorAfiveSet) \in \ColorSet$ where $\ColorAfiveSet$ does not contain $(N_d^{\highlightChange{r}}(X_i) = X_j)$,
    the size of the color class $(\ColorPartitionSet, \ColorAoneSet, \ColorAfiveSet)$ is 0.

    \item For every $(N_d^{\highlightChange{r}}(X_i) = X_j) \in A_5^-$,
    there exists a color $(\ColorPartitionSet, \ColorAoneSet, \ColorAfiveSet) \in \ColorSet$ such that $\ColorAfiveSet$ does not contain $(N_d^{\highlightChange{r}}(X_i) = X_j)$
    and the size of the color class $(\ColorPartitionSet, \ColorAoneSet, \ColorAfiveSet)$ is at least 1.

    \item For every $(|X_i| \leq m) \in A_6^+$,
    the sum of the sizes of all color classes $(\ColorPartitionSet, \ColorAoneSet, \ColorAfiveSet)$ with $i \in \ColorPartitionSet$ is at most~$m$.

    \item For every $(|X_i| \leq m) \in A_6^-$,
    the sum of the sizes of all color classes $(\ColorPartitionSet, \ColorAoneSet, \ColorAfiveSet)$ with $i \in \ColorPartitionSet$ is at least~$m+1$.

    \highlightChange{\item For every $\text{acy}(X_i) \in A_7^+$,
    the set of all vertices with colors $(\ColorPartitionSet, \ColorAoneSet, \ColorAfiveSet)$, where $i \in \ColorPartitionSet$, induces an acyclic subgraph.

    \item For every $\text{acy}(X_i) \in A_7^-$,
    the set of all vertices with colors $(\ColorPartitionSet, \ColorAoneSet, \ColorAfiveSet)$, where $i \in \ColorPartitionSet$, induces a non-acyclic subgraph.

    \item For every $\text{con}(X_i) \in A_8^+$,
    the set of all vertices with colors $(\ColorPartitionSet, \ColorAoneSet, \ColorAfiveSet)$, where $i \in \ColorPartitionSet$, induces a connected subgraph.

    \item For every $\text{con}(X_i) \in A_8^-$,
    the set of all vertices with colors $(\ColorPartitionSet, \ColorAoneSet, \ColorAfiveSet)$, where $i \in \ColorPartitionSet$, induces a non-connected subgraph.}
\end{itemize}

\highlightChange{In~\cite{DN-logic}, graphs are assumed to have an associated weight $w(v, S) \in \N$ for all vertices $v$ and sets $S \subseteq [k]$, and the weight of a $k$-tuple $B \in \mathcal{P}(V(G))^k$ is defined as $\sum_{v \in V(G)} w(v, \{i : v \in B_i\})$.}
Therefore, we can consider the weight set $(\WeightSet, \wlesseq, \weightsSum) = (\N \cup \{-\infty\}, \geq, +)$ and define the weight function as $\textsc{w}(v, (\ColorPartitionSet, \ColorAoneSet, \ColorAfiveSet), \highlightChange{\ell}) = w(v, \ColorPartitionSet)$ for every $v \in V(G)$, $(\ColorPartitionSet, \ColorAoneSet, \ColorAfiveSet) \in \ColorSet$ and \highlightChange{$\ell \in \N^{|\ColorSet| \times |R|}$}.

Now in order to compute an optimal solution of \highlightChange{$\Pi$}, we have to compute an optimal solution \highlightChange{of the locally checkable problem associated to} each truth assignment $\highlightChange{T}$ of the primitive formulas which satisfies $\psi$\highlightChange{, and then simply choose the best solution amongst them}.
As mentioned above, \highlightChange{the number of these truth assignments is at most $2^{\vert\psi\vert}$ and, by~\cite[Observation 3.1]{DN-logic}, this is at most $2^{10\vert\varphi\vert}$}.
\end{proof}

\subsection{Formulation of \texorpdfstring{$d$}{d}-stable locally checkable problems in DN logic}
\label{sec:LCPtoDN}
In this section we show
that any $d$-stable locally checkable problem \highlightChange{for $t$ distances} using $q$ colors, \highlightChange{size constraints $\sizeConstrSet$}, connectivity constraints $\mathcal{C}$ and acyclicity constraints $\mathcal{A}$, can be expressed as \highlightChange{formulas} in A\&C DN logic whose \highlightChange{lengths are} bounded by a function in $d$\highlightChange{,} $q$ \highlightChange{and $t$}.
This result, together with Theorem 1.2 in~\cite{DN-logic}, implies that such a problem is XP parameterized by
\highlightChange{the mim-width of a given binary decomposition tree of the input graph,}
when \highlightChange{all} $d$\highlightChange{,} $q$ \highlightChange{and $t$} are bounded by a constant.
Moreover, it allows us to combine locally checkable problems with connectivity and acyclicity constraints, as well as with any other property that can be potentially expressed in this logic.
However, the complexity of using this method turns out to be slightly worse than the one of the algorithm in \Cref{sec:Algorithm}, which, on the other hand, only allows connectivity constraints but not acyclicity.
Indeed, for a $d$-stable locally checkable problem with $q$ colors and no other constraints, the complexity given by the algorithm in \Cref{sec:Algorithm} is $n^{O(w q d)}$ while the complexity resulting from the next lemma is $n^{O(w q^2 d^2 (d+1)^{2q})}$, where $w$ is the mim-width of the given binary decomposition tree.
However, both results imply that the problems are XP parameterized by $w$.

\begin{lemma}\label{lemma:LCPtoDN}
Consider a locally checkable problem $\Pi$ on a graph $G$, with color set $\ColorSet = \set{a_1,\ldots,a_q}$, \highlightChange{check and weight functions that are $d$-stable for $t$ distances $r_1, \ldots, r_t$}, \highlightChange{size constraints $\sizeConstrSet \subseteq \partt_q(|V(G)|)$},
\highlightChange{positive and negative} connectivity constraints $\highlightChange{\mathcal{C}^{+}, \mathcal{C}^{-}} \subseteq \mathcal{P}([q])$,
and
\highlightChange{positive and negative} acyclicity constraints $\highlightChange{\mathcal{A}^{+}, \mathcal{A}^{-}} \subseteq \mathcal{P}([q])$.
Then\highlightChange{, there exists a set $\Phi=\{\varphi_1, \ldots, \varphi_{|\sizeConstrSet|}\}$ of A\&C DN formulas such that solving $\Pi$ is equivalent to finding the optimal among all the optimal solutions of the problems associated to the formulas in $\Phi$, and the length of each $\varphi\in\Phi$ is $O(t q^2 d (d+1)^{2qt})$.}
\end{lemma}
\begin{proof}
\highlightChange{Let $\sizeConstr \in \sizeConstrSet$.
We will show how to give an A\&C DN formula $\varphi_{\sizeConstr}$ that expresses the problem of finding a minimum weight proper coloring that satisfies the size constraint $\sizeConstr$ and the connectivity and acyclicity constraints $\mathcal{C}^{+}$, $\mathcal{C}^{-}$, $\mathcal{A}^{+}$ and $\mathcal{A}^{-}$.
The desired set $\Phi$ is then the set $\set{\varphi_{\sizeConstr}:\, \sizeConstr\in\sizeConstrSet}$.
Indeed, an optimal solution of $\Pi$ has to satisfy one of the size constraints of $\sizeConstrSet$.
An optimal solution of $\Pi$ must then be the optimal solution amongst a set of optimal solution for each size constraint in $\sizeConstrSet$.}

Let $\mathcal{T} = [q] \times \set{0,\ldots,d}^{q\highlightChange{t}}$.
For every $(j, \highlightChange{\kappa}) \in \mathcal{T}$\highlightChange{, with $j \in [q]$ and $\kappa = ((\kappa^1_1, \ldots, \kappa^1_q), \ldots, (\kappa^t_1, \ldots, \kappa^t_q)) \in \set{0,\ldots,d}^{qt}$}, we define the set $P_{(j, \highlightChange{\kappa})} = \set{v \in V(G) : check(v, a_j, \highlightChange{\kappa}) = \True}$.

First, notice that finding a minimum weight proper coloring is equivalent to finding a minimum weight partition of $V(G)$ into sets $X_T$, \highlightChange{for} $T \in \mathcal{T}$, such that the following conditions are satisfied:
\begin{enumerate}
\item\label{cond:proper} $check(v, a_j, \highlightChange{\kappa}) = \True$ (or, equivalently, $v \in P_{(j, \highlightChange{\kappa})}$) for all tuples $(j, \highlightChange{\kappa}) \in \mathcal{T}$ and $v \in X_{(j, \highlightChange{\kappa})}$,

\item\label{cond:neigh} $\highlightChange{\kappa^i_j} = \min\left(d, \left\vert N^{\highlightChange{r_i}}(v) \cap \left(\bigcup_{\highlightChange{\ell \in \set{0, \ldots, d}^{qt}}} X_{(j, \highlightChange{\ell})}\right)\right\vert\right)$ for \highlightChange{all $i \in [t]$,} all $j \in [q]$, all tuples $(j, \highlightChange{\kappa}) \in \mathcal{T}$ and \highlightChange{all} $v \in X_{(j, \highlightChange{\kappa})}$,

\item\label{cond:size} for all $j \in [q]$, the number of vertices in the set $\bigcup_{\highlightChange{\kappa \in \set{0, \ldots, d}^{qt}}} X_{(j, \highlightChange{\kappa})}$ equals $\sizeConstr_j$ (that is, the size constraint is satisfied for every color class),

\item\label{cond:conn} for all $C \in \mathcal{C}^{\highlightChange{+}}$ \highlightChange{(respectively $\mathcal{C}^{-}$)}, the graph induced by the set $\bigcup_{j \in C} \bigcup_{\highlightChange{\kappa \in \set{0, \ldots, d}^{qt}}} X_{(j, \highlightChange{\kappa})}$ is connected \highlightChange{(respectively not connected),} that is, the connectivity constraints are satisfied,

\item\label{cond:acyc} for all $A \in \mathcal{A}^{\highlightChange{+}}$ \highlightChange{(respectively $\mathcal{A}^{-}$)}, the graph induced by the set $\bigcup_{j \in A} \bigcup_{\highlightChange{\kappa \in \set{0, \ldots, d}^{qt}}} X_{(j, \highlightChange{\kappa})}$ is acyclic \highlightChange{(respectively not acyclic),} that is, the acyclicity constraints are satisfied,
\end{enumerate}
where the weights are defined as $w_{(j, \highlightChange{\kappa})}(v) = \textsc{w}(v, a_j, \highlightChange{\kappa})$.
In other words, we partition the vertices according to the color they receive and\highlightChange{, for each $i \in [q]$ and $r \in R$,} the number (up to $d$) of \highlightChange{vertices of color $a_i$ that they have at distance at most $r$.}

\highlightChange{In the next formulas we will make use of the following formula defined in~\cite[Proposition 6.4]{DN-logic}:
\[
\text{part}(X_{T \in \mathcal{T}}) \equiv \overline{\bigcup_{T\in\mathcal{T}}X_T}=\emptyset\land\bigwedge_{T,T'\in\mathcal{T}}X_T\cap X_{T'}=\emptyset.
\]
In other words, $\text{part}(X_{T \in \mathcal{T}})$ evaluates to $\True$ if the vertex sets $X_{T \in \mathcal{T}}$ partition the set $V(G)$.
We will, as well, employ the following function, which is defined for some positive integer $r$ and some finite or co-finite set $\mu\subseteq\N$:
\[
t_{r,\mu}(X) = \set{v\in V(G) : \vert N^r(v)\cap X\vert\in\mu}.
\]
This function was given in~\cite[Lemma 6.1]{DN-logic}, where it was also shown that $t_{r,\mu}(X)$ can be expressed as a DN term of length $O(d(\mu)x)$, where $x$ is the length of a term expressing $X$.}

Consider the following A\&C DN formula $\varphi_{\highlightChange{\sizeConstr}}$.
We claim that it expresses the existence of \highlightChange{a partition satisfying the above 5 conditions}.
\begin{align*}
\varphi_{\highlightChange{\sizeConstr}}(X_{T \in \mathcal{T}}) \equiv\;
&\text{part}(X_{T \in \mathcal{T}})
\land
\text{proper}(X_{T \in \mathcal{T}})
\land
\text{neigh}(X_{T \in \mathcal{T}})\\
&\land
\text{size}(X_{T \in \mathcal{T}})
\land
\text{conn}(X_{T \in \mathcal{T}})
\land
\text{acyc}(X_{T \in \mathcal{T}}),
\end{align*}
\highlightChange{where:
\begin{itemize}
\item {\color{black}$\text{proper}(X_{T \in \mathcal{T}}) \equiv
        \bigwedge_{T \in \mathcal{T}} X_T \subseteq P_T$},

\item {\color{black}$\text{neigh}(X_{T \in \mathcal{T}}) \equiv
        \bigwedge_{(j, \highlightChange{\kappa}) \in \mathcal{T}}
            X_{(j, \highlightChange{\kappa})}
            \subseteq
            \left(\highlightChange{\bigcap_{i \in [t]}} \bigcap_{h \in [q]} t_{\highlightChange{r_i}, f(\highlightChange{\kappa^i_h})}\left(\bigcup_{\highlightChange{\ell \in \set{0, \ldots, d}^{qt}}} X_{(h, \highlightChange{\ell})}\right)\right)$},
where {\color{black}the function $f$ is defined to be such that $f(x) = \set{x}$ if $x < d$ and $f(x) = \set{y \in \mathbb{N} : y\geq x}$ otherwise},

\item {\color{black}$\text{size}(X_{T \in \mathcal{T}}) \equiv
    \left(
        \bigwedge_{j \in [q]} \left\vert\bigcup_{\highlightChange{\kappa \in \set{0, \ldots, d}^{qt}}} X_{(j, \highlightChange{\kappa})} \right\vert = \sizeConstr_j
\right)$},

\item {\color{black}$\text{conn}(X_{T \in \mathcal{T}}) \equiv
        \left(\bigwedge_{C \in \mathcal{C}^{\highlightChange{+}}} \text{con}\left(\bigcup_{j \in C} \bigcup_{\highlightChange{\kappa \in \set{0, \ldots, d}^{qt}}} X_{(j, \highlightChange{\kappa})}\right)\right)
        \highlightChange{\land
        \left(\bigwedge_{C \in \mathcal{C}^-} \lnot\text{con}\left(\bigcup_{j \in C} \bigcup_{\kappa \in \set{0, \ldots, d}^{qt}} X_{(j, \kappa)}\right)\right)}$},

\item {\color{black}$\text{acyc}(X_{T \in \mathcal{T}}) \equiv
        \left(\bigwedge_{A \in \mathcal{A}^{\highlightChange{+}}} \text{acy}\left(\bigcup_{j \in A} \bigcup_{\highlightChange{\kappa \in \set{0, \ldots, d}^{qt}}} X_{(j, \highlightChange{\kappa})}\right)\right)
        \highlightChange{\land
        \left(\bigwedge_{A \in \mathcal{A}^-} \lnot\text{acy}\left(\bigcup_{j \in A} \bigcup_{\highlightChange{\kappa \in \set{0, \ldots, d}^{qt}}} X_{(j, \highlightChange{\kappa})}\right)\right)}$}.
\end{itemize}}

By definition, $\varphi_{\highlightChange{\sizeConstr}}$ expresses the existence of sets $X_{T \in \mathcal{T}}$ such that each of the subformulas $\text{part}(X_{T \in \mathcal{T}})$, $\text{proper}(X_{T \in \mathcal{T}})$, $\text{neigh}(X_{T \in \mathcal{T}})$, $\text{size}(X_{T \in \mathcal{T}})$, $\text{conn}(X_{T \in \mathcal{T}})$ and $\text{acyc}(X_{T \in \mathcal{T}})$ holds.
As mentioned above, $\text{part}(X_{T \in \mathcal{T}})$ guarantees that the sets $X_{T \in \mathcal{T}}$ form a partition of $V(G)$.
The subformula $\text{proper}(X_{T \in \mathcal{T}})$ expresses condition~\ref{cond:proper} for all the vertices.
\highlightChange{By definition of $t_{r,\mu}$}, we have that the subformula $\text{neigh}(X_{T \in \mathcal{T}})$ assures that \highlightChange{for every $(j, \kappa) \in \mathcal{T}$ we have $X_{(j, \kappa)} \subseteq \bigcap_{i \in [t]} \bigcap_{h \in [q]} \{v \in V(G) : \left\vert N^{r_i}(v) \cap \left(\bigcup_{\ell \in \set{0, \ldots, d}^{qt}} X_{(h, \ell)}\right)\right\vert \in f(\kappa^i_h)\}$.
In other words, it assures that} every vertex $v \in X_{(j, \highlightChange{\kappa})}$ is such that $\left\vert N^{\highlightChange{r_i}}(v) \cap \left(\bigcup_{\highlightChange{\ell \in \set{0, \ldots, d}^{qt}}} X_{(h, \highlightChange{\ell})}\right)\right\vert \in f(\highlightChange{\kappa^i_h})$ for \highlightChange{every $i \in [t]$ and} every $h \in [q]$, which implies condition~\ref{cond:neigh}.
Finally, $\text{size}(X_{T \in \mathcal{T}})$, $\text{conn}(X_{T \in \mathcal{T}})$ and $\text{acyc}(X_{T \in \mathcal{T}})$ ensure that the size, connectivity and acyclicity constraints are satisfied (conditions~\ref{cond:size} to~\ref{cond:acyc}).

\highlightChange{Notice that $\mathcal{T}$ has $q(d+1)^{qt}$ elements.
Then, t}he length of the formula $\text{part}(X_{T\in\mathcal{T}})$ is $O(q^2 (d+1)^{2q\highlightChange{t}})$ and the length of $\text{proper}(X_{T \in \mathcal{T}})$ is $O(q(d+1)^{q\highlightChange{t}})$.
Since the length of a term $t_{\highlightChange{r_i}, f(\highlightChange{\kappa^i_h})}(X)$ is $O(d \cdot \text{length}(X))$, the length of $\text{neigh}(X_{T \in \mathcal{T}})$ is $O(\highlightChange{t} q^2 d (d+1)^{2q\highlightChange{t}})$.
The lengths of $\text{size}(X_{T\in\mathcal{T}})$, $\text{conn}(X_{T\in\mathcal{T}})$ and $\text{acyc}(X_{T\in\mathcal{T}})$ are, respectively, $O(q (d+1)^{\highlightChange{qt}})$, $O(|\mathcal{C}| q (d+1)^{\highlightChange{qt}})$ and $O(|\mathcal{A}| q (d+1)^{\highlightChange{qt}})$\highlightChange{, since there are at most $q(d+1)^{qt}$ possible tuples $(j, \kappa)$ and for the last two formulas we have a conjunction of $|\mathcal{C}|$ and $|\mathcal{A}|$ terms of length $O(q (d+1)^{qt})$ each.}
Therefore, \highlightChange{and since $|\mathcal{C}| \leq 2^{q}$ and $|\mathcal{A}| \leq 2^{q}$,} the length of the complete formula is \highlightChange{$O(t q^2 d (d+1)^{2qt})$}.
\end{proof}

By~\cite[Theorem 1.2]{DN-logic}, \highlightChange{each of the problems associated to the formulas $\varphi_1, \ldots, \varphi_{|\sizeConstrSet|}$ of a problem $\Pi$ as in \Cref{lemma:LCPtoDN}} can be solved in time
\highlightChange{$n^{O(w t^2 q^4 d^3 (d+1)^{4qt})}$}
where $w$ is the mim-width of the given binary decomposition tree.
Since $\vert\sizeConstrSet\vert \leq n^{\highlightChange{q-1}}$, we do not have to apply the algorithm more than $n^{\highlightChange{q-1}}$ times.
We can then look for an optimal solution among the $\vert\sizeConstrSet\vert$ solutions we have obtained.
\highlightChange{Hence, the total complexity of solving the problem $\Pi$ is $n^{O(w t^2 q^4 d^3 (d+1)^{4qt})}$.}

\section{Solving \texorpdfstring{$d$}{d}-stable locally checkable problems}
\label{sec:Algorithm}
In this section we give a more straightforward algorithm to solve $d$-stable locally checkable problems with size and connectivity constraints.
Its complexity is slightly better than the one obtained with the DN logic formulation.
\highlightChange{To be precise, we will show the following theorem:}
\highlightChange{
\mainResult*
}

\highlightChange{We} first define some necessary terms and notation.
Let $d$ be a positive integer.
We define an equivalence relation $=_d$ on the natural numbers, where two non-negative integers $i$ and $j$ are equivalent if and only if $\min(d,i) = \min(d,j)$.
For a binary decomposition tree $(T, \delta)$ of a graph $G$ and a node $v \in V(T)$, 
\highlightChange{we make use of the equivalence relation $\equiv_d^v$ presented in~\cite{LCVSVP-paper2}.
It is defined} on subsets of $T_v$ where two subsets $S, S' \subseteq T_v$ are equivalent if and only if for any $y \in T_{\overline{v}}$ we have $\vert N(y) \cap S\vert =_d \vert N(y) \cap S'\vert$.
\highlightChange{Intuitively, from a perspective of locally checkable problems, we could say that two subsets of $T_v$ are equivalent if replacing one for the other would not influence the output of any $d$-stable check function when applied to any vertex $y$ in $T_{\overline{v}}$.
The idea behind this is that we do not want to save all partial solutions in a dynamic programming algorithm, and (roughly speaking) it is enough to save one representative for each equivalence class.
One of the key ideas when working with this equivalence class is then to give an upper bound for the number of equivalence classes which then leads to an upper bound of the number of partial solutions that we have to save.
Such an upper bound in terms of the mim-width of the decomposition of the graph was given in \cite{GraphClassesWithStructuredNeighborhoods}, where it was shown that the number of equivalence classes of $\equiv_d^v$ is at most $O(n^{d\cdot mimw(G)})$.}

\highlightChange{Based on $\equiv_d^v$, we}
define the equivalence relation $\equiv_{d,q}^v$ on $q$-tuples of subsets of $T_v$, where two $q$-tuples $X, X' \in (\mathcal{P}(T_v))^q$ are equivalent if and only if they are entry-wise equivalent according to $\equiv_d^v$.
Analogously, we define the equivalence relations $\equiv_d^{\overline{v}}$ and $\equiv_{d,q}^{\overline{v}}$ where subsets $S, S'$ of $T_{\overline{v}}$ are equivalent if for every vertex $w$ in $T_v$ we have $\vert N(w) \cap S\vert =_d \vert N(w) \cap S'\vert$, and analogously for tuples in $(\mathcal{P}(T_{\overline{v}}))^q$.
We will mainly use these relations when referring to partitions of $T_v$ and not arbitrary tuples of subsets.

\highlightChange{For an} equivalence relation $\equiv$ on a set $S$ and $X \in S$, we denote by $[X]^{\highlightChange{\equiv}}$ the equivalence class which consists of all elements $X' \in S$ such that $X' \equiv X$. 
\highlightChange{When the equivalence relation is clear from context, we will drop the superscript and only write $[X]$.}
We denote by $\mathcal{R}_{d}^v$ and $\mathcal{R}_{d,q}^v$ the set of all equivalence classes of the relations $\equiv_{d}^v$ and $\equiv_{d,q}^{v}$, respectively, and analogously for $\overline{v}$.
If $R \in \mathcal{R}_{d,q}^v$ and $X \in (\mathcal{P}(T_{v}))^q$ we write $X \equiv_{d,q}^v R$ if $X$ is contained in the equivalence class $R$ and say that $X$ is a \highlightChange{\emph{representative}} of $R$ (and analogously for $\overline{v}$).
If $R \in \mathcal{R}_d^v$ and $w \in T_{\overline{v}}$ we define $\vert N(w) \cap R\vert$ to be the unique number $i \in \set{0, \ldots, d}$ such that $i =_d \vert N(w) \cap S\vert$ for some $S \subseteq T_v$ with $S \equiv_d^v R$.
This notion is well-defined by the definition of $\equiv_d^v$.
We say that $w$ is \emph{adjacent} to $R$ if $\vert N(w)\cap R\vert\geq 1$ and we say that a set $S\subseteq T_{\overline{v}}$ is adjacent to $R$ if it contains a vertex which is adjacent to $R$.
If $C \subseteq [q]$ and $X \in (\mathcal{P}(T_v))^q$, we denote by $X_C$ the set $\bigcup_{i\in C}X_i \subseteq T_v$ and we also define the equivalence class $[X]_C$ of the relation $\equiv_d^v$ by setting $[X]_C = [X_C]$.
We do everything analogously for $T_{\overline{v}}$.
We define $\snec{d}(T) = \max_{v \in V(T)}(\max\{\vert\mathcal{R}_{d}^v\vert, \vert\mathcal{R}_{d}^{\overline{v}}\vert\})$ and $\snec{d,q}(T) = \max_{v \in V(T)}(\max\{\vert\mathcal{R}_{d,q}^v\vert, \vert\mathcal{R}_{d,q}^{\overline{v}}\vert\})$.

\begin{remark}\label{remark:rep}
It was shown in~\cite[Lemma 1]{LCVSVP-paper2} that for a node $v$ of $T$ we can compute a list which contains exactly one representative of each equivalence class of $\equiv_{d}^{v}$, in time $O(\snec{d}(T) \log(\snec{d}(T)) |V(G)|^2)$.
The same applies to $\equiv_{d}^{\overline{v}}$.
\highlightChange{By~\cite[Lemma 2]{GraphClassesWithStructuredNeighborhoods}, $\snec{d}(T) \leq n^{d \cdot mimw(T,\delta)}$.}
Since the equivalence classes of $\equiv_{d,q}^{v}$ and $\equiv_{d,q}^{\overline{v}}$ are $q$-tuples of equivalence classes of $\equiv_{d}^{v}$ and $\equiv_{d}^{\overline{v}}$, respectively, their representatives are simply the $q$-tuples of the representatives of $\equiv_{d}^{v}$ and $\equiv_{d}^{\overline{v}}$, respectively.
\highlightChange{Thus, we can also efficiently compute a list of representatives and $\snec{d,q}(T) \leq n^{q \cdot d \cdot mimw(T,\delta)}$.}
\end{remark}

The next \highlightChange{lemma} shows that it is possible to solve, in polynomial time for graphs of bounded mim-width with a given suitable binary decomposition tree, a locally checkable problem with a $d$-stable check function, a $d$-stable weight function, a set of \highlightChange{positive} connectivity constraints $\mathcal{C} \subseteq \mathcal{P}([q])$ and a size constraint $\sizeConstrSet$ with $|\sizeConstrSet| = 1$.
We will later explain how this result can be easily extended to a set $\sizeConstrSet$ of several size constraints \highlightChange{and to negative connectivity constraints}.
For the complexity analysis, we will use the fact that $d$ \highlightChange{and} $q$ are constants and that $\vert\C\vert \leq 2^q$ (and hence is bounded by a constant).

\begin{lemma}
\label{thm:mainThm}
Let $d, q$ be integers, with $q \geq 2$.
Consider a $d$-stable locally checkable problem $\Pi$ with $q$ colors, a \highlightChange{set $\sizeConstrSet$ containing only one size constraint $\sizeConstr$,} and \highlightChange{a set of positive} connectivity constraints $\mathcal{C}$.
Then, for an input graph $G$ with an associated binary decomposition tree $(T, \delta)$, we can solve $\Pi$ in
$O\left(
    n^{\highlightChange{2q}}
    \cdot
    \snec{d,q}(T)^{6 \vert\C\vert + 6}
    \cdot
    \left(
        n^2
        +
        \snec{d,q}(T)^{\omega - 1}
    \right)
\right)$
time,
where $n = |V(G)|$ and $\omega$ is the matrix multiplication exponent ($\omega<2.4$,~\cite{BodlaenderCyganKratschNederlof}).
\end{lemma}
\begin{proof}
This proof closely follows the ones given in~\cite{LCVSVP-paper2} and~\cite{bergougnoux2021applications}.
Our focus will be on highlighting that the $d$-stability of the check function is enough to guarantee the correctness of the output, and how the weight function and the restriction on the size of color classes can be implemented.

First of all, notice that finding a coloring $c$ of $G$ with $q$ colors $\{a_1, \ldots, a_q\}$ is equivalent to finding a $q$-partition $(X_1, \ldots, X_q)$ of $V(G)$, where $X_j = c^{-1}(a_j)$ for all $j \in [q]$.
We will therefore show how to compute a $q$-partition $X \in \partt_q(V(G))$ such that
$check(v, a, |N(v) \cap X_1|, \ldots, |N(v) \cap X_q|) = \True$ for all $v \in V(G)$ and $a \in \ColorSet$,
$|X| = \sizeConstr$ and
$G[X_C]$ is connected for all $C \in \mathcal{C}$.
\highlightChange{We will assume that $G$ does not contain isolated vertices (if it did, we could solve the locally checkable problem independently in each connected component).}

For every $v \in V(T)$, $X \in \partt_q(T_v)$ and $R' \in \mathcal{R}_{d,q}^{\overline{v}}$ and some $Y \in \partt_q(T_{\overline{v}})$ with $Y \equiv_{d,q}^{\overline{v}} R'$ we define the weight of $X$ when complemented with \highlightChange{$Y$, $\textsc{w}(X,Y)$, as}
\[
\highlightChange{\textsc{w}(X,Y)}
=
\bigweightsSum_{j \in [q]}
\bigweightsSum_{w \in X_j}
    \textsc{w}(
        w,
        a_j,
        \vert N(w) \cap (X_1 \cup Y_1)\vert,
        \ldots,
        \vert N(w) \cap (X_q \cup Y_q)\vert
    )
.
\]
\highlightChange{We define $\textsc{w}(X,[Y]) = \textsc{w}(X,Y)$.
It follows from the $d$-stability of $\textsc{w}$ and the definition of the equivalence relation $\equiv_{d,q}^{\overline{v}}$ that this notion is well-defined.}

We say that a tuple $(X,Y) \in \partt_q(T_v) \times \partt_q(T_{\overline{v}})$ is \emph{\TvalidY{v}} if for any vertex $w \in T_v$ we have $check(w, j_w, \vert N(w) \cap (X_1 \cup Y_1)\vert, \ldots, \vert N(w) \cap (X_q \cup Y_q)\vert) = \True$, where $j_w$ is such that $w \in X_{j_w}$, $\vert X\vert + \vert Y\vert = \sizeConstr$ and $G[X_C \cup Y_C]$ is connected for each $C \in \mathcal{C}$.
We say that a tuple $(X,R') \in \partt_q(T_v) \times \mathcal{R}_{d,q}^{\overline{v}}$ is \emph{\TvalidR{v}} if for any vertex $w \in T_v$ and for a representative $Y \equiv_{d,q}^{\overline{v}} R'$ we have $check(w, j_w, \vert N(w) \cap (X_1 \cup Y_1)\vert, \ldots, \vert N(w) \cap (X_q \cup Y_q)\vert) = \True$, where $j_w$ is such that $w \in X_{j_w}$.
Observe that all of these notions are well-defined by the definition of the equivalence relation $\equiv_{d,q}^v$.

We now explain the general idea of the algorithm.
For every node $v\in V(T)$ and for every $R\in \mathcal{R}_{d,q}^v$, $R'\in\mathcal{R}_{d,q}^{\overline{v}}$, $\sizeConstr' \in \partt_q(\vert T_v\vert)$, we will compute a set $M_v[R,R',\sizeConstr']\subseteq \partt_q(T_v)$ such that
\highlightChange{\begin{itemize}
\item {\color{black}$\vert M_v[R,R',\sizeConstr']\vert\leq 2^{\vert\C\vert}\snec{d,q}(T)^{2\vert\C\vert}$,}
\item {\color{black}for every $X\in M_v(R,R',\sizeConstr')$ we have that $X\equiv_{d,q}^v R$, $\vert X\vert = \sizeConstr'$ and that $(X,R')$ is {\TvalidR{v}}}, and
\item {\color{black} for every $X\in\partt_q(T_v)$ with $X\equiv_{d,q}^v R$, $\vert X\vert = \sizeConstr'$ and every $Y\in\partt_q(T_{\overline{v}})$ such that $Y\equiv_{d,q}^{\overline{v}} R'$ and $(X,Y)$ is {\TvalidY{v}} there is an $X'\in M_v[R,R',\sizeConstr']$ such that $\textsc{w}(X',Y)\wlesseq \textsc{w}(X,Y)$ and $(X',Y)$ is {\TvalidY{v}}.}
\end{itemize}}

Let $r$ be the root of $T$.
Observe that $T_{\overline{r}}=\emptyset$ and thus for every $X,X'\in \partt_q(T_r)$, we have that $X \equiv_{d,q}^{r} X'$.
We further have that $\R_{d,q}^r$ contains only one element, namely $[\emptyset^q]$.
It follows that $M_r[[\emptyset^q],[\emptyset^q],\sizeConstr]$ has to contain an optimal solution of our problem, if and only if such a solution exists.
In order to solve the problem we just need to search for an $X \in M_r[[\emptyset^q], [\emptyset^q], \sizeConstr]$ of minimum weight $\textsc{w}(X, \emptyset^{q})$ such that $(X, \emptyset^{q})$ is {\TvalidY{r}}.
\highlightChange{Notice that the size of $M_r[[\emptyset^q], [\emptyset^q], \sizeConstr]$ is at most $2^{\vert\C\vert}\snec{d,q}(T)^{2\vert\C\vert}$, and that checking if $(X, \emptyset^{q})$ is {\TvalidY{r}} (knowing that $(X, [\emptyset^q])$ is \TvalidR{r}) and computing $\textsc{w}(X, \emptyset^{q})$ can be done in $O(n^2)$ time.
Therefore, given the set $M_r[[\emptyset^q],[\emptyset^q],\sizeConstr]$, finding an optimal solution there (or discovering that no solution exists)} can be done in time $O(\snec{d,q}(T)^{2\vert\C\vert} n^2)$.

We will compute these sets $M_v[R,R',\sizeConstr']$ by the standard dynamic programming bottom-up approach.
We start by computing them whenever $v$ is a leaf of $T$.
Then, $T_v$ contains a single vertex $w$ of $G$ and for each equivalence class in $\mathcal{R}_{d,q}^v$ there is just one representative.
Thus, for each $R\in\mathcal{R}_{d,q}^v$ and $R'\in\mathcal{R}_{d,q}^{\overline{v}}$ we let $X$ be the unique representative of $R$ and if $X\in\partt_q(T_v)$ and $(X, R')$ is {\TvalidR{v}} then we set $M_v[R,R',\vert X\vert] = \set{X}$ and to $\emptyset$ if not.
For every $\sizeConstr' \neq \vert X\vert$, we set $M_v[R,R',\sizeConstr'] = \emptyset$.

Now let $v$ be a non-leaf node of $T$.
Let $a$ and $b$ be the children of $v$, and assume we have already correctly computed the sets $M_a[R_a, R'_a, \sizeConstr_a]$ for every $R_a \in \mathcal{R}_{d,q}^a$, $R'_a \in \mathcal{R}_{d,q}^{\overline{a}}$ and $\sizeConstr_a \in \partt_q(\vert T_a\vert)$, as well as $M_b[R_b, R'_b, \sizeConstr_b]$ for every $R_b \in \mathcal{R}_{d,q}^b$, $R'_b \in \mathcal{R}_{d,q}^{\overline{b}}$ and $\sizeConstr_b \in \partt_q(\vert T_b\vert)$.
For $R \in \mathcal{R}_{d,q}^v$, $R' \in \mathcal{R}_{d,q}^{\overline{v}}$ and $\sizeConstr' \in \partt_q(\vert T_v\vert)$, we say that two tuples $(R_a,R_a',\sizeConstr^a)\in\mathcal{R}_{d,q}^a \times \mathcal{R}_{d,q}^{\overline{a}} \times \partt_q(\vert T_a\vert)$ and $(R_b,R_b',\sizeConstr^b)\in\mathcal{R}_{d,q}^b \times \mathcal{R}_{d,q}^{\overline{b}}\times\partt_q(\vert T_b\vert)$ are \emph{$(R,R',\sizeConstr')$-compatible} if
$\sizeConstr^a + \sizeConstr^b = \sizeConstr'$
and, for representatives $Y, Y', Y_a, Y'_a, Y_b, Y'_b$ of $R, R', R_a, R'_a, R_b, R'_b$ respectively, we have
$Y_a \cup Y_b \equiv_{d,q}^v Y$, $Y_b \cup Y' \equiv_{d,q}^{\overline{a}} Y'_a$, $Y_a \cup Y' \equiv_{d,q}^{\overline{b}} Y'_b$.
For some fixed $R \in \mathcal{R}_{d,q}^v$, $R' \in \mathcal{R}_{d,q}^{\overline{v}}$ and $\sizeConstr' \in \partt_q(\vert T_v\vert)$ consider the set 
\[
\mathcal{M} =
\bigcup_{\substack{(R_a, R_a', \sizeConstr^a) \textrm{ and } (R_b, R_b', \sizeConstr^b)\\
\textrm{are } (R,R',\sizeConstr')-\textrm{compatible}}
}
\left(\bigcup_{\substack{X_a\in M_a[R_a,R_a',\sizeConstr^a]\\X_b\in M_b[R_b,R_b',\sizeConstr^b]}} (X_a\cup X_b)\right).
\]
Here the union is indexed over all tuples $(R_a, R_a', \sizeConstr^a)$ and $(R_b, R_b', \sizeConstr^b)$ which are as above and $(R,R',\sizeConstr')$-compatible.
We first show that this set has all the properties demanded from $M_v[R,R',\sizeConstr']$ except for the upper bound on the size\highlightChange{, and t}hen we show a way to reduce the size of the set without losing these other properties.
\highlightChange{Notice that the size of $\mathcal{M}$ could be as large as $n^q \cdot \snec{d,q}(T)^4 \cdot \left(2^{\vert\C\vert}\snec{d,q}(T)^{2\vert\C\vert}\right)^2$ because we are assuming that $M_a[R_a,R_a',\sizeConstr^a]$ and $M_b[R_b,R_b',\sizeConstr^b]$ are of size at most $2^{\vert\C\vert}\snec{d,q}(T)^{2\vert\C\vert}$ and there are at most $n^q \cdot \snec{d,q}(T)^4$ pairs of tuples $(R_a, R_a', \sizeConstr^a)$ and $(R_b, R_b', \sizeConstr^b)$ that are $(R,R',\sizeConstr')$-compatible (there are at most $n^q$ possible values for $\sizeConstr^a$ and for each of them there is only one choice for $\sizeConstr^b$, and there are at most $\snec{d,q}(T)$ possible values for each $R_a, R_a', R_b, R_b'$).}

Let $(R, R', \sizeConstr')$ be as above and $X \in \partt_q(T_v)$ be such that $X \equiv_{d,q}^v R$ and $|X| = \sizeConstr'$.
Suppose there exists $Y \in \partt_q(T_{\overline{v}})$ such that $Y \equiv_{d,q}^{\overline{v}} R'$ and $(X,Y)$ is {\TvalidY{v}}.
Set $X^a = T_a \cap X$, $X^b = T_b \cap X$, $X'^a = X^b \cup Y$, $X'^b = X^a \cup Y$, $\sizeConstr^a = |X^a|$ and $\sizeConstr^b = |X^b|$.
Clearly $(X^a,X'^a)$ is {\TvalidY{a}}, $(X^b,X'^b)$ is {\TvalidY{b}} and $\sizeConstr^a + \sizeConstr^b = \sizeConstr'$, hence $M_a[[X^a],[X'^a],\sizeConstr^a]$ and $M_b[[X^b],[X'^b],\sizeConstr^b]$ are not empty.

Let $\Conn(Z^a, Z^b, Y, \mathcal{C})$ denote the statement ``$(Z^a)_C \cup (Z^b)_C \cup Y_C$ is connected for each $C \in \mathcal{C}$''.
Observe that
\begin{align*}
\textsc{w}(X, Y)
&= \textsc{w}(X^a \cup X^b, Y)\\
&= \textsc{w}(X^a, X'^a) \weightsSum \textsc{w}(X^b, X'^b)\\
&\wgteq\min
    _{\substack{
        Z^a \in \partt_q(T_a),\; Z^a \equiv_{d,q}^a X^a\\
        Z^b \in \partt_q(T_b),\; Z^b \equiv_{d,q}^b X^b\\
        \vert Z^a\vert = K^a,\; \vert Z^b\vert = K^b\\
        \Conn(Z^a, Z^b, Y, \mathcal{C})
    }}
    \left(
        \textsc{w}(Z^a, X'^a) \weightsSum \textsc{w}(Z^b, X'^b)
    \right)\\
&=
\min
    _{\substack{
        Z^a \in \partt_q(T_a),\; Z^a \equiv_{d,q}^a X^a\\
        \vert Z^a\vert = K^a
    }}
    \left(
        \textsc{w}(Z^a, X'^a)
        \weightsSum
        \left(\min
                _{\substack{
                    Z^b \in \partt_q(T_b),\; Z^b \equiv_{d,q}^b X^b\\
                    \vert Z^b\vert = K^b\\
                    \Conn(Z^a, Z^b, Y, \mathcal{C})
                }}
                \left(
                    \textsc{w}(Z^b, X'^b)
                \right)
        \right)
    \right)\\
&=\min
    _{\substack{
        Z^a \in \partt_q(T_a),\; Z^a \equiv_{d,q}^a X^a\\
        \vert Z^a\vert = K^a
    }}
    \left(
        \textsc{w}(Z^a, X'^a)
        \weightsSum
        \left(\min
                _{\substack{
                    Z^b \in M_b[[X^b],[X'^b],\sizeConstr^b]\\
                    \Conn(Z^a, Z^b, Y, \mathcal{C})
                }}
                \left(
                    \textsc{w}(Z^b, X'^b)
                \right)
        \right)
    \right)\\
&=\min
    _{\substack{
        Z^a \in \partt_q(T_a),\; Z^a \equiv_{d,q}^a X^a\\
        \vert Z^a\vert = K^a\\
        Z^b \in M_b[[X^b],[X'^b],\sizeConstr^b]\\
        \Conn(Z^a, Z^b, Y, \mathcal{C})
    }}
    \left(
        \textsc{w}(Z^a, X'^a) \weightsSum \textsc{w}(Z^b, X'^b)
    \right)\\
&=\min
    _{\substack{
        Z^b \in M_b[[X^b],[X'^b],\sizeConstr^b]
    }}
    \left(
        \left(\min
                _{\substack{
                    Z^a \in \partt_q(T_a),\; Z^a \equiv_{d,q}^a X^a\\
                    \vert Z^a\vert = K^a\\
                    \Conn(Z^a, Z^b, Y, \mathcal{C})
                }}
                \left(
                    \textsc{w}(Z^a, X'^a)
                \right)
        \right)
        \weightsSum
        \textsc{w}(Z^b, X'^b)
    \right)\\
&=\min
    _{\substack{
        Z^b \in M_b[[X^b],[X'^b],\sizeConstr^b]
    }}
    \left(
        \left(\min
                _{\substack{
                    Z^a \in M_a[[X^a],[X'^a],\sizeConstr^a]\\
                    \Conn(Z^a, Z^b, Y, \mathcal{C})
                }}
                \left(
                    \textsc{w}(Z^a, X'^a)
                \right)
        \right)
        \weightsSum
        \textsc{w}(Z^b, X'^b)
    \right)\\
&=\min
    _{\substack{
        Z^a \in M_a[[X^a],[X'^a],\sizeConstr^a]\\
        Z^b \in M_b[[X^b],[X'^b],\sizeConstr^b]\\
        \Conn(Z^a, Z^b, Y, \mathcal{C})
    }}
    \left(
        \textsc{w}(Z^a, X'^a) \weightsSum \textsc{w}(Z^b, X'^b)
    \right)\\
&=\min
    _{\substack{
        Z^a \in M_a[[X^a],[X'^a],\sizeConstr^a]\\
        Z^b \in M_b[[X^b],[X'^b],\sizeConstr^b]\\
        \Conn(Z^a, Z^b, Y, \mathcal{C})
    }}
    \left(
        \textsc{w}(Z^a,Z^b \cup Y) \weightsSum \textsc{w}(Z^b, Z^a \cup Y)
    \right)\\
&=\min
    _{\substack{
        Z^a \in M_a[[X^a],[X'^a],\sizeConstr^a]\\
        Z^b \in M_b[[X^b],[X'^b],\sizeConstr^b]\\
        \Conn(Z^a, Z^b, Y, \mathcal{C})
    }}
    \textsc{w}(Z^a \cup Z^b, Y)\highlightChange{.}
\end{align*}

It follows that there are $Z^a \in M_a[[X^a],[X'^a],\sizeConstr^a]$ and $Z^b \in M_b[[X^b],[X'^b],\sizeConstr^b]$ such that $\textsc{w}(Z^a \cup Z^b, Y) \wlesseq \textsc{w}(X^a \cup X^b, Y) = \textsc{w}(X,Y)$.
Further, $(Z^a\cup Z^b,Y)$ is {\TvalidY{v}}.
Indeed, the connectivity constraints are satisfied by the above and the other constraints follow from the definition of $M_a$ and $M_b$.

Succinctly, for every $X \in \partt_q(T_v)$ with $X \equiv_{d,q}^v R$ and $|X| = \sizeConstr'$, and every $Y \in \partt_q(T_{\overline{v}})$ such that $Y \equiv_{d,q}^{\overline{v}} R'$ and $(X,Y)$ is {\TvalidY{v}}\highlightChange{,} there exists $Z \in \mathcal{M}$ such that $\textsc{w}(Z, Y) \wlesseq \textsc{w}(X,Y)$ and $(Z,Y)$ is {\TvalidY{v}}.

\highlightChange{W}e are now going to show that we can find a subset of $\mathcal{M}$ which has cardinality at most $2^{\vert\C\vert}\snec{d,q}(T)^{2\vert\C\vert}$ and keeps the property we have just shown.
Observe that any subset of $\mathcal{M}$ can only contain partitions of $T_v$ which are {\TvalidR{v}} together with $R'$ and who have the correct partition size.
Thus, when choosing a subset of $\mathcal{M}$, we only have to focus on keeping the \highlightChange{partial solutions which are part of an optimal solution}.

Let $v \in V(T)$, $R' \in \mathcal{R}_{d,q}^{\overline{v}}$, $\mathcal{X} \subseteq \partt_q(T_v)$ as well as $\mathcal{X}' \subseteq \mathcal{X}$.
We say that $\mathcal{X}'$ does \emph{$(v,R')$-represent} $\mathcal{X}$ if, for every
\highlightChange{$X \in \mathcal{X}$ and $Y \in \partt_q(T_{\overline{v}})$ such that $Y \equiv_{d,q}^{\overline{v}} R'$ and} $X_C \cup Y_C$ is connected for every $C \in \mathcal{C}$, there exists $X' \in \mathcal{X}'$ such that $X'_C \cup Y_C$ is connected for every $C \in \mathcal{C}$ and $\textsc{w}(X',R') \wlesseq \textsc{w}(X,R')$.

\begin{claim}\label{claim}
Let $v\in V(T)$ and $R\in\mathcal{R}_{d,q}^v$, $R'\in\mathcal{R}_{d,q}^{\overline{v}}$.
Let $\mathcal{X}\subseteq\partt_q(T_v)$ be such that $X\equiv_{d,q}^v R$ for every $X\in\mathcal{X}$.
We can find a set $(v,R')$-representing $\mathcal{X}$ of cardinality at most $2^{\vert\C\vert}\snec{d,q}(T)^{2\vert\C\vert}$ in
$O(
    |\mathcal{X}|
    \cdot
    \snec{d,q}(T)^{2\vert\C\vert}
    \cdot
    (n^2 + \snec{d,q}(T)^{\omega - 1})
)$
time,
where $\omega$ is the matrix multiplication exponent ($\omega<2.4$~\cite{BodlaenderCyganKratschNederlof}).
\end{claim}
\begin{claimproof}
Let $\highlightChange{\mathcal{C}_{\emptyset}} \subseteq\mathcal{C}$ be the set of all connectivity restraints $C$ for which $\emptyset \equiv_{d}^{\overline{v}} R'_C$.
For any $C \in \highlightChange{\mathcal{C}_{\emptyset}}$ and for any $Y \in \partt_q(T_{\overline{v}})$ with $Y\equiv_{d,q}^{\overline{v}} R'$, notice that $Y_C \equiv_d^{\overline{v}} \emptyset$.
Also, if $Y_C = \emptyset$ then $Y$ can only be completed to a {\TvalidY{v}} coloring with a partition $X \in \mathcal{X}$ when $X_C$ is connected; on the other hand, if $Y_C\neq\emptyset$ then $Y$ can only be completed to a {\TvalidY{v}} coloring with a partition $X \in \mathcal{X}$ when $X_C$ is empty, since otherwise $X_C \cup Y_C$ cannot be connected.

Given a map $\tau: \highlightChange{\mathcal{C}_{\emptyset}} \rightarrow\set{0,1}$ we say that a partition $X\in\mathcal{X}$ is of \emph{type $\tau$} if $X_C=\emptyset$ for every $C\in\tau^{-1}(0)$ and $X_C$ is connected for every $C\in\tau^{-1}(1)$.
Observe that if $\highlightChange{\mathcal{C}_{\emptyset}} = \emptyset$ then $\tau$ can only be the empty map and then every $X$ is of type $\tau$.
There are at most $2^{\vert\highlightChange{\mathcal{C}_{\emptyset}}\vert}$ possible maps from $\highlightChange{\mathcal{C}_{\emptyset}}$ to $\set{0,1}$.
For each map $\tau$ as above we will compute a set $\mathfrak{X}_{\tau}$ of size at most $\snec{d,q}(T)^{2\vert\C\vert}$ that $(v,R')$-represents the set $\set{X \in \mathcal{X} : X\textrm{ has type }\tau}$.
We claim that their union will yield a set as demanded.
Indeed, for every $Y\equiv_{d,q}^{\overline{v}}R'$ we define a map $\tau_Y$ as above by mapping $C \in \highlightChange{\mathcal{C}_{\emptyset}}$ to 0 if $Y_C$ is non-empty and to 1 otherwise.
By the above, for any $X \in \mathcal{X}$ such that $X_C \cup Y_C$ is connected for every $C \in \mathcal{C}$,
we have that $X$ is of type $\tau_Y$.
This implies that there is $X' \in \mathfrak{X}_{\tau_{Y}}$ such that $X'_C \cup Y_C$ is connected for every $C \in \mathcal{C}$
and $\textsc{w}(X',Y) \wlesseq \textsc{w}(X,Y)$.
Hence the union of all the $\mathfrak{X}_{\tau}$ is a $(v,R')$-representative of $\mathcal{X}$.
Further, if the size of each $\mathfrak{X}_{\tau}$ is indeed at most $\snec{d,q}(T)^{2\vert\C\vert}$, then their union has size at most $2^{\vert\highlightChange{\mathcal{C}_{\emptyset}}\vert}\snec{d,q}(T)^{2\vert\C\vert}\leq 2^{\vert\C\vert}\snec{d,q}(T)^{2\vert\C\vert}$.

It remains to show how to compute such a set $\mathfrak{X}_{\tau}$ for a fixed $\tau$.
Fix a map $\tau$ as above and consider the set $\mathcal{X}_{\tau}$ consisting of every $X \in \mathcal{X}$ which has type $\tau$.
Let $\highlightChange{\overline{\mathcal{C}_{\emptyset}}} = \mathcal{C} \setminus \highlightChange{\mathcal{C}_{\emptyset}}$.
Now $R'_C$ does not contain the empty set for any $C \in \highlightChange{\overline{\mathcal{C}_{\emptyset}}}$.
Let $\mathcal{Y}$ be the set of all partitions $Y \equiv_{d,q}^{\overline{v}} R'$ such that $\tau_Y = \tau$.

If there is a partition $X \in \mathcal{X}_{\tau}$ such that there is a $C \in \highlightChange{\overline{\mathcal{C}_{\emptyset}}}$ and a connected component $\connectedComp\subseteq X_C$ such that $\connectedComp$ is not adjacent to $R'_C$, then for any $Y \in \mathcal{Y}$ the set $X_C \cup Y_C$ is not connected.
Hence, we remove all such partitions from $\mathcal{X}_{\tau}$.
Analogously, if there is a partition $Y \in \mathcal{Y}$ which has a connected component $\connectedComp$ of $Y_C$ such that $\connectedComp$ is not adjacent to $R_C$ then $X_C \cup Y_C$ cannot be connected for any $X \in \mathcal{X}_{\tau}$.
Hence, we remove these partitions from $\mathcal{Y}$.

If the size of $\mathcal{X}_{\tau}$ is at most $\snec{d,q}(T)^{2\vert\C\vert}$, then we can set $\mathfrak{X}_{\tau}=\mathcal{X}_{\tau}$.
Thus, we can assume from now on that $\vert\mathcal{X}_{\tau}\vert >\snec{d,q}(T)^{2\vert\C\vert}$.

For every $S\subseteq T_{\overline{v}}$ let $v_S$ be a vertex in $S$.
Let $\T$ be the set of $\highlightChange{\overline{\mathcal{C}_{\emptyset}}}$-tuples whose entries are 2-tuples from $\R_{d,q}^{\overline{v}}$
(that is, $\T$ is the set of all elements $(R_C^1,R_C^2)_{C \in \highlightChange{\overline{\mathcal{C}_{\emptyset}}}}$ with $R^1, R^2 \in \R_{d,q}^{\overline{v}}$).
Let us now define the following matrices over $\mathbb{F}_2$:
\begin{itemize}
\item $J$ is the $(\mathcal{X}_{\tau} \times \mathcal{Y})$-matrix for which $J(X,Y)=1$ if and only if $X_C\cup Y_C$ is connected for each $C \in \highlightChange{\overline{\mathcal{C}_{\emptyset}}}$.

\item $L$ is a $(\mathcal{P}(T_v) \times (\R_{d}^{\overline{v}} \times \R_{d}^{\overline{v}}))$-matrix such that $L(X,(R'_1,R'_2))=1$ if and only if no connected component of $X$ is adjacent to both $R'_1$ and $R'_2$.

\item $\overline{L}$ is a $((\R_{d}^{\overline{v}} \times \R_{d}^{\overline{v}}) \times \mathcal{P}(T_{\overline{v}}))$-matrix such that $\overline{L}((R'_1,R'_2),Y)=1$ if and only if there is a connected cut $(Y_1,Y_2)$ of $Y$ such that $v_Y\in Y_1$, $Y_1\equiv_{d}^{\overline{v}} R'_1$ and $Y_2\equiv_{d}^{\overline{v}} R'_2$.

\item $L^*$ is an $(\mathcal{X}_{\tau}\times\T)$-matrix such that $L^*(X,(R_C^1,R_C^2)_{C\in\C}) = 1$ if and only if $L(X_C,(R_C^1,R_C^2))=1$ for each $C \in \highlightChange{\overline{\mathcal{C}_{\emptyset}}}$.

\item $\overline{L^*}$ is a $(\T\times \mathcal{Y})$-matrix such that $\overline{L^*}((R_C^1,R_C^2)_{C\in\C},Y)=1$ if and only if $\overline{L}((R_C^1,R_C^2),Y_C) = 1$ for every $C \in \highlightChange{\overline{\mathcal{C}_{\emptyset}}}$.
\end{itemize}

\highlightChange{We will now prove} that $L^*\overline{L^*} = J$.
\highlightChange{The operations that follow are all over $\mathbb{F}_2$.}
\highlightChange{For all} $X\in\mathcal{X}_{\tau}$ \highlightChange{and all} $ Y\in\mathcal{Y}$, we have\highlightChange{:}

\begin{align*}
L^*\overline{L^*}(X,Y)
&=
\sum
    _{(R_C^1, R_C^2)_{C \in \highlightChange{\overline{\mathcal{C}_{\emptyset}}}} \in \T}
    \mathbbm{1}\left(
        \forallFormula
            {C \in \highlightChange{\overline{\mathcal{C}_{\emptyset}}}}
            {L\left(X_C, \left(R_C^1, R_C^2\right) \right) = 1
            \land
            \overline{L}\left( \left(R_C^1, R_C^2\right), Y_C\right) = 1}
    \right)\\
&=
\vert \{
    (R_C^1, R_C^2)_{C \in \highlightChange{\overline{\mathcal{C}_{\emptyset}}}} \in \T
    \,:\,
    L\left(X_C, \left(R_C^1,R_C^2\right) \right) = 1
    \land
    \overline{L}\left( \left(R_C^1,R_C^2\right), Y_C\right) = 1\\
&\qquad\qquad
        \textrm{for every } C \in \highlightChange{\overline{\mathcal{C}_{\emptyset}}}
\} \vert\\
&=
\vert \{
    (R_C^1, R_C^2)_{C \in \highlightChange{\overline{\mathcal{C}_{\emptyset}}}} \in \T
    \,:\,
    \forallFormula
        {C \in \highlightChange{\overline{\mathcal{C}_{\emptyset}}}}
        {\existsFormula
            {
                (Z_C^1, Z_C^2) \in ccut(X_C \cup Y_C)
            }
            {}}\\
&\qquad\qquad
                v_{Y_C} \in Z_C^1
                \land
                Z_C^1 \cap Y_C \equiv_{d}^{\overline{v}} R_C^1
                \land
                Z_C^2 \cap Y_C \equiv_{d}^{\overline{v}} R_C^2
\} \vert\\
&=
    \prod
        _{C \in \highlightChange{\overline{\mathcal{C}_{\emptyset}}}}
        \vert \{
            (R^1,R^2) \in \R_{d}^{\overline{v}} \times \R_{d}^{\overline{v}}
            \,:\,
            \existsFormula
                {
                    (Z^1, Z^2) \in ccut(X_C \cup Y_C)
                }
                {}\\
&\qquad\qquad
                    v_{Y_C} \in Z^1
                    \land
                    Z^1 \cap Y_C \equiv_{d}^{\overline{v}} R^1
                    \land
                    Z^2 \cap Y_C \equiv_{d}^{\overline{v}} R^2
        \} \vert\\
&=
\prod
    _{C \in \highlightChange{\overline{\mathcal{C}_{\emptyset}}}}
    \left\vert \left\{
        (Z^1, Z^2) \in ccut(X_C \cup Y_C)
        \,:\,
        v_{Y_C} \in Z^1
    \right\} \right\vert\\
&=
\prod
    _{C \in \highlightChange{\overline{\mathcal{C}_{\emptyset}}}}
    2^{\vert cc(X_C \cup Y_C) \vert - 1}
.
\end{align*}

The first two equalities follow right from the definition, the third one follows from~\cite[Claim 4.3.1]{bergougnoux2021applications}.
The fourth one holds since we can choose the pair of equivalence classes $(R_C^1, R_C^2)$ independently for each $C \in \highlightChange{\overline{\mathcal{C}_{\emptyset}}}$.
The fifth equality follows from~\cite[Claim 4.3.2]{bergougnoux2021applications}, and the last one follows from the fact that any connected component of $X_C\cup Y_C$ which does not contain $v_{Y_C}$ can arbitrarily be assigned to both $Z^1$ and $Z^2$.
Since we defined our matrices to be over $\F_2$, and the only case where every factor of the last product is odd is when $X_C \cup Y_C$ is connected for every $C \in \highlightChange{\overline{\mathcal{C}_{\emptyset}}}$, then the claim follows.

Observe that we can compute the matrix $L^*$ in $O(|\mathcal{X}_{\tau}| \cdot \snec{d,q}(T)^{2\vert\C\vert} \cdot n^2)$ time.
Indeed, the number of entries of $L^*$ is at most $|\mathcal{X}_{\tau}| \cdot \snec{d,q}(T)^{2\vert\C\vert}$.
Also, for an entry indexed by $(X, (R_C^1,R_C^2)_{C\in\C})$ and for every $C \in \highlightChange{\overline{\mathcal{C}_{\emptyset}}}$, we can first compute the connected components of $X_C$ and then check the adjacencies.
All these steps can be done in $O(n^2)$ time.

Now, we assign each row of $L^*$ a weight.
More precisely, the row which is indexed by $X\in\mathcal{X}_{\tau}$ obtains the weight $\textsc{w}(X,R')$.
These assignments can be done in $O(\vert\mathcal{X}_{\tau}\vert n^2)$ time.
Let $\B \subseteq \mathcal{X}_{\tau}$ be such that the rows of $L^*$ which are indexed by the entries in $\B$ are a basis of the row-space of $L^*$, $row(L^*)$, and the sum of the weights of these rows is minimum among all bases of $row(L^*)$ which consist of row vectors of $L^*$.
Observe that the dimension of $row(L^*)$ and thus the size of \highlightChange{$\B$} is at most the number of columns of $L^*$, which is bounded from above by $\snec{d,q}(T)^{2\vert\C\vert}$.
Observe as well that, by~\cite[Lemma 3.15]{BodlaenderCyganKratschNederlof}, $\B$ can be computed in time $O(|\mathcal{X}_{\tau}| \cdot \snec{d,q}(T)^{2\vert\C\vert(\omega - 1)})$ (where $\omega<2.373$ is the matrix multiplication coefficient).

We claim that $\B$ is a set which has the properties demanded from $\mathfrak{X}_{\tau}$.
The size is already correct, then it remains to show that for every $X\in\mathcal{X}_{\tau}$, $Y \in \mathcal{Y}$, with $J(X,Y) = 1$, there is an $X' \in \B$ such that $J(X',Y) = 1$ and $\textsc{w}(X',Y) \wlesseq \textsc{w}(X,Y)$.
Indeed, since $\B$ corresponds to a basis over $\F_2$, there is a unique $\B' \subseteq \B$ such that for any $\rho \in \T$ we have
\[
    L^*(X,\rho) = \sum_{B\in\B'}L^*(B,\rho).
\]

We can use this to rewrite the row of $J$ corresponding to $X$:
\begin{align*}
    J(X,Y)
            &= \sum_{\rho \in \T} L^*(X, \rho) \overline{L^*}(\rho, Y)\\
            &= \sum_{B \in \B'} \sum_{\rho \in \T} L^*(B, \rho) \overline{L^*}(\rho, Y)\\
            &= \sum_{B \in \B'} J(B, Y).
\end{align*}
Hence, if $J(X,Y) = 1$ then there is an odd number of elements $B\in\B'$ such that $J(B,Y) = 1$.
Let $B$ be any of them which has maximum weight.
If $\textsc{w}(X,Y) \wless \textsc{w}(B,Y)$ then $\B \setminus \set{B} \cup \set{X}$ corresponds to a basis of the row-space of $L^*$ which has smaller weight than the one corresponding to $\B$.
Indeed, it is clear that the weight would be smaller by the assumption on its weight and it does indeed correspond to a basis, since $L^*(B,\rho) = L^*(X,\rho) - \sum_{B' \in \B \setminus \set{B}} L^*(B',\rho)$ for every $\rho \in \T$.
\end{claimproof}

\highlightChange{It follows from \Cref{claim} that there is a set of size at most $2^{\vert\C\vert}\snec{d,q}(T)^{2\vert\C\vert}$ which $(v,R')$-represents $\mathcal{M}$.
We can thus set $M_v[R,R',\sizeConstr']$ to be any such set.}
It remains to determine the runtime of the entire algorithm.

We can precalculate a representative for each equivalence class using \Cref{remark:rep}.
For a fixed node $v \in V(T)$ the matrix $M_v$ has at most $\snec{d,q}(T)^2 \cdot n^{\highlightChange{q-1}}$ entries.
We compute those entries using dynamic programming in a bottom-up fashion.
When $v$ is a leaf of $T$, each of those sets can be computed in constant time.
When $v$ is not a leaf, we first compute $\mathcal{M}$ in $O\left(n^q \cdot \snec{d,q}(T)^4 \cdot \left(2^{\vert\C\vert} \cdot \snec{d,q}(T)^{2\vert\C\vert}\right)^2 \cdot n^2\right)$ time by using the pre-calculated sets over $a$ and $b$ \highlightChange{(for each of the $O(n^q \cdot \snec{d,q}(T)^4)$ pairs of compatible tuples, we iterate over all possible $X_a, X_b$, which are at most $\left(2^{\vert\C\vert} \cdot \snec{d,q}(T)^{2\vert\C\vert}\right)^2$ possibilities, and compute $X_a \cup X_b$ in $O(n^2)$ time)}, and then we apply \Cref{claim} in
\highlightChange{$O(n^q \cdot \snec{d,q}(T)^4 \cdot (2^{\vert\C\vert} \cdot \snec{d,q}(T)^{2\vert\C\vert})^2 \cdot \snec{d,q}(T)^{2\vert\C\vert} \cdot (n^2 + \snec{d,q}(T)^{\omega - 1}))$
time, since in this case we have $|\mathcal{X}| = |\mathcal{M}| = n^q \cdot \snec{d,q}(T)^4 \cdot (2^{\vert\C\vert} \cdot \snec{d,q}(T)^{2\vert\C\vert})^2$}.
It follows that every entry of $M_v$ can be computed in time
$O\left(
    n^q
    \cdot
    \snec{d,q}(T)^{6 \vert\C\vert + 4}
    \cdot
    \left(
        n^2
        +
        \snec{d,q}(T)^{\omega - 1}
    \right)
\right)$.
\highlightChange{Then we only need to multiply by the possible number of nodes $v \in V(T)$, which is $O(n)$, and by the number of possible entries of $M_v$, which is $O(\snec{d,q}(T)^2 \cdot n^{q-1})$.
We conclude by a}dding the complexity of the analysis at the root\highlightChange{.
Therefore}, we obtain the claimed total runtime.
\end{proof}

In this proof, we only considered the case when we want one fixed size for the sizes of the partition sets.
However, this is enough to handle \emph{any} restrictions on the sizes of the partition sets.
Indeed, let $G$ be a graph with $n$ vertices together with a check function $check$ and a weight function $\textsc{w}$ and some connectivity restraints $\mathcal{C}\subseteq \mathcal{P}([q])$.
If we want to find an optimal $q$-partition of $V(G)$ which satisfies the constraints given by $check$ and $\mathcal{C}$ for which the size of the partition classes is contained in some $\sizeConstrSet \subseteq \partt_q(n)$ then we can determine an optimal partition for each $\sizeConstr \in \sizeConstrSet$ with the algorithm above.
Since $\vert \sizeConstrSet \vert \leq n^{\highlightChange{q-1}}$, we do not have to apply the algorithm more than a polynomial number of times. 
We can then look for an optimal solution among the $\vert \sizeConstrSet \vert$ solutions we have determined.

\highlightChange{In order to handle a set $\C^-$ of negative connectivity constraints, first notice that asking for a set $S$ to be disconnected is equivalent to asking that $S$ can be partitioned into two non-empty subsets $S_1$ and $S_2$ such that there are no edges between vertices in $S_1$ and vertices in $S_2$.
Hence, we can do the following: replace the set of colors $\ColorSet$ with the set $\ColorSet\times\left(\bigtimes_{C\in\C^{-}}\set{0,1}\right)$.
This can be seen as keeping the old colors but adding a flag for every negative connectivity constraint.
We refer to the flag associated to some $C\in\C^-$ as the $C$-flag.
The check and weight functions maintain their functionality while ignoring the newly introduced flags.
However, if a vertex $v$ receives color $a$ (in the first entry) and this color is contained in some $C\in\C^-$, the check function now also makes sure that $v$ has no neighbor with a color $a'\in C$ in the first entry and with a different $C$-flag.
If a vertex $v$ receives a color $a$ (in the first entry) that is not contained in some $C\in\C^-$, its $C$-flag has no importance whatsoever.
We now ask further that for every $C\in\C^-$ there is at least one vertex whose color is in $C$ and whose $C$-flag is 0 and that there is at least one vertex whose color is in $C$ and whose $C$-flag is 1.
This last constraint is a size constraint on the color classes.
The new number of colors is $q 2^{|\C^-|}$, which is not larger than $q 2^{2^{q}}$.}

Since $\snec{d,q}(T) \leq n^{q d \highlightChange{w}}$\highlightChange{, where $w$ is the mim-width of the given decomposition} (see \highlightChange{\Cref{remark:rep}}), we can say that \highlightChange{if we do not have negative connectivity constraints, then} the time complexity of the algorithm is
$O\left(
    n^{\highlightChange{2q} + \highlightChange{w} d q (6 \vert\C\vert + \omega + 5)}
\right)$, which can also be written as \highlightChange{$n^{O(w d q (\vert\C\vert + 1))}$}.
\highlightChange{However, in the case where we have a set $\C^-$ of negative connectivity constraints and a set $\C^+$ of positive ones, then the complexity is
$O\left(
    n^{2 q 2^{|\C^-|} + w d q 2^{|\C^-|} (6 \vert\C^+\vert + \omega + 5)}
\right)$, which can also be written as $n^{O(w d q (\vert\C^+\vert + 1) 2^{\C^-})}$.}
\highlightChange{Hence, our main result (\Cref{thm:mainResult}) follows.}

\subsection{An extension to distance versions}
\label{sec:Distance}
In the setup we have just seen, the check and weight functions obtain as input only information about the colors of the vertices in the neighborhood of $v$.
Similarly to the approach in \cite{LCPTreewidth}, we can extend this to \highlightChange{larger distances}.

Let $t$ be a fixed positive integer.
Let $r_1, \ldots, r_t$ be positive integers.
It is possible to modify the algorithm above to handle \highlightChange{check and weight functions which are $d$-stable for $t$ distances $r_1, \ldots, r_t$.}
We will only sketch the outlines of how to do this.
We know from \cite{gen-dist-dom-mim-width-2019} that for any graph $G$, positive integer $r$ and binary decomposition tree $(T, \delta)$ of $G$, the mim-width of $(T, \delta)$ can at most double when it is applied to $G^r$.
We define $\equiv_d^{v,r}$ and $\equiv_{d,q}^{v,r}$ as before, where now $r$ specifies the graph to which we refer, $G^r$.
Analogously, define $\R_d^{v,r}$ and $\R_{d,q}^{v,r}$, as well as all the above for $\overline{v}$.
We also define $T_v^r$-partial validity for a tuple $(X, R'^1, \ldots, R'^t)$ by only considering representatives $Y \subseteq \partt_q(T_{\overline{v}})$ for which $Y \equiv_{d,q}^{\overline{v},r_i} R'^i$ for every $i \in [t]$.
For every $v\in V(T)$, $R^i \in \R_{d,q}^{v,r_i}$, $R'^i \in \R_{d,q}^{\overline{v},r_i}$, with $i \in [t]$, and $K' \in \partt_q(\vert T_v\vert)$ we compute $M_v[R^1, R'^1, \ldots, R^t, R'^t, K'] \subseteq \partt_q(T_v)$ with the conditions as before and additionally for every $X$ in this set we demand that $X \equiv_{d,q}^{v,r_i} R^i$ for every $i \in [t]$.
The size constraint for $M_v$ will remain the same.
The rest of the algorithm will not be changed fundamentally.
Observe in particular that \Cref{claim} can be used unchanged, since we would only have \emph{more} restrictions to put on $Y$.
When computing $\mathcal{M}$, we have to consider more possibilities of combinations, but after the reduction step, the size will be the same as in the 1-distance version, only the computation time changes.

Let us explain the new time complexity.
We first obtain the graphs $G^{r_1}, \ldots, G^{r_t}$ and the corresponding representatives.
Each of the matrices $M_v$ have $\snec{d,q}(T)^{2t} n^q$ entries.
The number of elements in each of them is still at most $2^{\vert\C\vert} \cdot \snec{d,q}(T)^{2\vert\C\vert}$.
The base case can still be done in constant time (since we are assuming that $t$ is a constant).
When $v$ is not a leaf, we compute $\mathcal{M}$ and reduce it with \Cref{claim} in
$O(n^q \cdot \snec{d,q}(T)^{4t} \cdot (2^{\vert\C\vert} \cdot \snec{d,q}(T)^{2\vert\C\vert})^2 \cdot (n^2 + \snec{d,q}(T)^{2\vert\C\vert} \cdot (n^2 + \snec{d,q}(T)^{\omega - 1})))$ time.
The overall time complexity is then
$O\left(
    n^{q+2}
    \cdot
    \snec{d,q}(T)^{6 \vert\C\vert + 6t}
    \cdot
    \left(
        n^2
        +
        \snec{d,q}(T)^{\omega - 1}
    \right)
\right)$.
Using the inequality on $\snec{d,q}(T)$, we obtain $n^{O(w d q (\vert\C\vert + t))}$.
\highlightChange{Observe that we can handle negative connectivity constraints in the same way as in \Cref{thm:mainResult}.
The following corollary then follows.}

\highlightChange{\begin{corollary}
\label{thm:mainResultDistances}
Let $d, q$ and $t$ be positive integers.
For a locally checkable problem with $q$ colors, a check function and a weight function that are $d$-stable for $t$ distances, size constraints $\sizeConstrSet$, positive connectivity constraints $\C^+$ and negative connectivity constraints $\C^-$, there exists an algorithm that solves the problem on graphs with a given binary decomposition tree in time
$n^{O(w d q (\vert\C^+\vert + t) 2^{\vert\C^-\vert})}$,
where $n$ is the number of vertices of the input graph and $w$ is the mim-width of the associated decomposition.
\end{corollary}}

We would like to highlight that only the number of different distances employed affects the complexity, but not the specific values of the distances.
This allows us to consider problems where, for example, each vertex of color 1 has to be at distance at most $\frac{n}{2}$ of a vertex of color 2.

\section{Limits of the framework}
\label{sec:Limits}
We have shown in the previous section that a $d$-stable locally checkable problem with a bounded number of colors can be solved on a family of graphs of bounded mim-width in polynomial time.
In this section we show that it is unlikely that this result can be extended to locally checkable problems with less restrictions.
In particular, we analyze here the following three situations: unbounded number of colors, weight functions which are not $d$-stable for any constant $d$, and also check functions which are not $d$-stable for any constant $d$.

\subsection{Non-constant number of colors}
First we show that it is unlikely that we can handle an unbounded number of colors, even when both the check and weight functions are $1$-stable.
\highlightChange{As noted in~\cite{LCPTreewidth}, obtaining the chromatic number of a graph $G$ can be modeled as a locally checkable problem with 1-stable functions but a linear number of colors:
\begin{itemize}
\item $\ColorSet = \{1, \ldots, |V(G)|\}$,

\item $(\WeightSet, \wlesseq, \weightsSum) = (\mathbb{N} \cup \{+\infty\}, \leq, \max)$,

\item $\textsc{w}(v, a, k_1, \ldots, k_{|V(G)|}) = a$
for all $v \in V(G)$, $a \in \ColorSet$ and $k_1, \ldots, k_{|V(G)|} \in \N$,

\item $check(v, a, k_1, \ldots, k_{|V(G)|}) = \left(k_a = 0\right)$
for every $v \in V(G)$, $a \in \ColorSet$ and $k_1, \ldots, k_{|V(G)|} \in \N$.
\end{itemize}

However, this problem was shown to be NP-complete on circular-arc graphs~\cite{ColoringNPhCircArc}, a graph class of mim-width at most 2~\cite{BELMONTE201354} (this fact was also noted in~\cite{mimwHereditaryGraphs}).}
Consequently, if we were able to find a method which could find optimal solutions of 1-stable locally checkable problems with unbounded number of colors in graphs of mim-width \highlightChange{2} in polynomial time, then we could solve any NP-complete problem in polynomial time.

\subsection{Non-\texorpdfstring{$d$}{d}-stable \highlightChange{color-counting} weight function}
\highlightChange{In this section we consider the Max-Cut problem.
Given a graph $G$, a \emph{cut} is a partition of the vertices into two disjoint sets $S$ and $T$, and its size is defined as the number of edges of the induced bipartite graph $G[S,T]$.
The Max-Cut problem asks for the maximum size of a cut in a graph.
This problem was shown to be NP-hard on interval graphs~\cite{MaxCutInterval}, which all have mim-width 1 and for which a decomposition of mim-width 1 can be computed in linear time \cite{GraphClassesWithStructuredNeighborhoods}.}
We can model \highlightChange{Max-Cut} as a locally checkable problem with two colors \highlightChange{and color-counting functions}.
Indeed, let
\begin{itemize}
\item $\ColorSet = \set{\textsc{s}, \textsc{t}}$,
\item $(\WeightSet, \wlesseq, \weightsSum) = (\mathbb{Q} \cup \{-\infty\}, \geq, +)$,
\item $\textsc{w}(v, \textsc{s}, k_\textsc{s}, k_\textsc{t}) = \frac{k_\textsc{t}}{2}$ and $\textsc{w}(v, \textsc{t}, k_\textsc{s}, k_\textsc{t}) = \frac{k_\textsc{s}}{2}$ for all $v \in V(G)$, $k_\textsc{s}, k_\textsc{t} \in \N$, and
\item $check(v, a, k_\textsc{s}, k_\textsc{t}) = \True$ for all $v \in V(G)$, $a \in \set{\textsc{s}, \textsc{t}}$, $k_\textsc{s}, k_\textsc{t} \in \N$.
\end{itemize}
In other words, every coloring is proper, and for a given coloring, the sum of all the weights is exactly the number of edges whose endpoints are colored differently.
Thus, a coloring of minimum weight (according to the order defined on the weight set) does indeed correspond to a maximum cut.
Observe that its check function is 1-stable, but the weight function is not $d$-stable for any constant $d$.
Thus, we cannot expect to eliminate this restriction on the weight function.

\newcommand{\gadgetClause}[3]{
    \draw[ultra thick] (#1, #2) -- (#1 + 13, #2);
        
    \draw[ultra thick] (#1 + 0.5, #2 - 0.5) -- (#1 + 4, #2 - 0.5);
    \draw[ultra thick] (#1 + 1, #2 - 1) -- (#1 + 2, #2 - 1);
    \draw[ultra thick] (#1 + 2.75, #2 - 1) -- (#1 + 3.75, #2 - 1);
        
    \draw[ultra thick] (#1 + 4.75, #2 - 0.5) -- (#1 + 8.25, #2 - 0.5);
    \draw[ultra thick] (#1 + 5.25, #2 - 1) -- (#1 + 6.25, #2 - 1);
    \draw[ultra thick] (#1 + 7, #2 - 1) -- (#1 + 8, #2 - 1);
        
    \draw[ultra thick] (#1 + 9, #2 - 0.5) -- (#1 + 12.5, #2 - 0.5);
    \draw[ultra thick] (#1 + 9.5, #2 - 1) -- (#1 + 10.5, #2 - 1);
    \draw[ultra thick] (#1 + 11.25, #2 - 1) -- (#1 + 12.25, #2 - 1);

    \def\temp{#3}\ifx\temp\empty
    \else
        \path (#1 - 0.25, #2) node{$v_{#3}$};
        \path (#1 + 0.25, #2 - 0.5) node{$x_{#3}$};
        \path (#1 + 0.75, #2 - 1) node{$x_{#3}'$};
        \path (#1 + 2.5, #2 - 1) node{$x_{#3}''$};
        \path (#1 + 4.5, #2 - 0.5) node{$y_{#3}$};
        \path (#1 + 5, #2 - 1) node{$y_{#3}'$};
        \path (#1 + 6.75, #2 - 1) node{$y_{#3}''$};
        \path (#1 + 8.75, #2 - 0.5) node{$z_{#3}$};
        \path (#1 + 9.25, #2 - 1) node{$z_{#3}'$};
        \path (#1 + 11, #2 - 1) node{$z_{#3}''$};%
    \fi
}

\newcommand{\gadgetLeft}[3]{
    \draw[ultra thick] (#1, #2) -- (#1 + 1.5, #2);
    \draw[ultra thick] (#1 + 1, #2 - 0.5) -- (#1 + 3.5, #2 - 0.5);
    \draw[ultra thick] (#1 + 2.75, #2 - 1) -- (#1 + 5, #2 - 1);
    \draw[ultra thick] (#1 + 2.75, #2 - 1.5) -- (#1 + 6.5, #2 - 1.5);

    \def\temp{#3}\ifx\temp\empty
    \else
        \path (#1 - 0.25, #2) node{$v_1$};
        \path (#1 + 0.75, #2 - 0.5) node{$v_2$};
        \path (#1 + 2.5, #2 - 1) node{$v_3$};
        \path (#1 + 2.5, #2 - 1.5) node{$v_4$};%
    \fi
}

\newcommand{\gadgetLeftSmall}[3]{
    \draw[ultra thick] (#1, #2) -- (#1 + 1.5, #2);
    \draw[ultra thick] (#1 + 1, #2 - 0.5) -- (#1 + 3, #2 - 0.5);
    \draw[ultra thick] (#1 + 1, #2 - 1) -- (#1 + 4.5, #2 - 1);

    \def\temp{#3}\ifx\temp\empty
    \else
        \path (#1 - 0.25, #2) node{$v_1$};
        \path (#1 + 0.75, #2 - 0.5) node{$v_2$};
        \path (#1 + 0.75, #2 - 1) node{$v_3$};
    \fi
}

\subsection{Non-\texorpdfstring{$d$}{d}-stable \highlightChange{color-counting} check function}
Finally, we show that for general color-counting check functions, we cannot expect to test existence of proper colorings unless P=NP.
\highlightChange{To this end, we define the following ad hoc locally checkable problem with constant number of colors and a color-counting check function.}
Let us set $\ColorSet=\set{1, \ldots, 7, \texttt{T}, \texttt{t}, \texttt{F}, \texttt{f}}$.
Given a vertex $v$, a color $a$ and the number $k_{a'}$ of neighbors of $v$ of color $a'$, for every $a' \in \ColorSet$, the check function
\highlightChange{is defined as follows:
\begin{align*}
check&(v, a, k_1, \ldots, k_7, k_{\texttt{T}}, k_{\texttt{t}}, k_{\texttt{F}}, k_{\texttt{f}}) = \\
          & \;(a=1 \Leftrightarrow \text{deg}(v) = 2) \\
    \land & \;(a=2 \Leftrightarrow \text{deg}(v) > 2 \land k_1 \geq 1) \\
    \land & \;(a=3 \Leftrightarrow \text{deg}(v) > 2 \land k_1 = 0 \land k_2 \geq 2) \\
    \land & \;(a=4 \Leftrightarrow \text{deg}(v) > 2 \land k_1 = 0 \land k_2 = 1) \\
    \land & \;(a=5 \Leftrightarrow \text{deg}(v) > 2 \land k_1 = 0 \land k_2 = 0 \land k_3 = k_4) \\
    \land & \;(a=6 \Leftrightarrow \text{deg}(v) > 2 \land k_1 = 0 \land k_2 = 0 \land k_3 \neq k_4 \land k_5 > 1) \\
    \land & \;(a=7 \Leftrightarrow \text{deg}(v) > 2 \land k_1 = 0 \land k_2 = 0 \land k_3 \neq k_4 \land k_5 = 1 \land \textstyle\sum_{i \in \set{7, \texttt{T}, \texttt{t}, \texttt{F}, \texttt{f}}} k_i = 2) \\
    \land & \;(a=5 \Rightarrow k_{\texttt{T}} = 1) \\
    \land & \;(a=6 \Rightarrow k_{\texttt{T}} = k_{\texttt{t}} \land k_{\texttt{F}} = k_{\texttt{f}}) \\
    \land & \;(a=7 \Rightarrow k_{\texttt{t}} = k_{\texttt{T}} = 1 \lor k_{\texttt{f}} = k_{\texttt{F}} = 1).
\end{align*}

Notice that it is justified to use the degree of the input vertex in the check function since the degree equals the sum of the sizes of all color classes in the neighborhood.}

We claim that it is NP-hard to determine the existence of a proper coloring even when restricted to interval graphs (and thus in a graph class where the mim-width is bounded by 1).
We show this by polynomially reducing from Positive 1-in-3 SAT, which is shown to be NP-complete in~\cite{tippenhauer}.
It is a 3-SAT variant in which all literals are positive and a truth assignment is satisfying if there is exactly one true literal in every clause.

\begin{figure}[t]
    \centering

    \begin{tikzpicture}
        \gadgetClause{0}{0}{c};
    \end{tikzpicture}

    \caption{Clause gadget for $c = (x, y, z)$.}
    \label{fig:clauseGadget}
\end{figure}
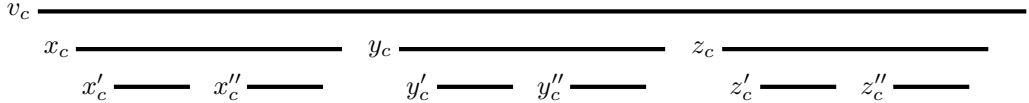

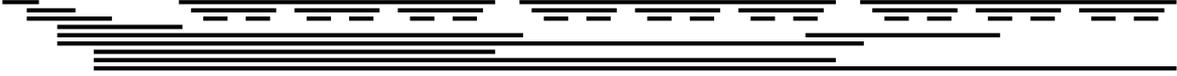
\begin{figure}[t]
    \centering

    \begin{tikzpicture}[xscale=0.32, yscale=0.22]
        \gadgetLeftSmall{-7.25}{0}{};

        \gadgetClause{0}{0}{};
        \gadgetClause{14}{0}{};
        \gadgetClause{28}{0}{};

        \draw[ultra thick] (-5, -1.5) -- (0.15, -1.5);
        \draw[ultra thick] (-5, -2) -- (14.15, -2);
        \draw[ultra thick] (-5, -2.5) -- (28.15, -2.5);

        \draw[ultra thick] (-3.5, -3) -- (13, -3);
        \draw[ultra thick] (-3.5, -3.5) -- (27, -3.5);
        \draw[ultra thick] (-3.5, -4) -- (41, -4);

        \draw[ultra thick] (25.75, -2) -- (33.75, -2);
    \end{tikzpicture}

    \caption{Example of the interval representation of $G_I$, where $I$ consists of three clauses $c_1 = (x_1, x_2, x_3)$, $c_2 = (x_4, x_5, x_6)$ and $c_3 = (x_7, x_6, x_8)$.}
    \label{fig:example}
\end{figure}

Given an instance $I$ (with at least one clause) of positive 1-in-3-SAT with clause set $C$ and variable set $X$, we construct an interval graph $G_I$ as follows.
For each clause $c = (x, y, z)$ we create a copy of the \emph{clause gadget} illustrated in \Cref{fig:clauseGadget}.
We call $v_c$ a \emph{clause vertex}, $x_c$, $y_c$ and $z_c$ \emph{variable vertices}, and $x_c'$, $x_c''$, $y_c'$, $y_c''$, $z_c'$ and $z_c''$ \emph{value vertices}.
All the clause gadgets are placed next to each other in an arbitrary order such that they do not overlap.
For every variable $x\in X$, do the following.
\highlightChange{Consider every two clauses $c$ and $c'$ that both contain $x$ and whose gadgets are positioned in such a way that no other gadget corresponding to a clause containing $x$ is positioned between the gadgets of $c$ and $c'$.}
Without loss of generality, assume that the gadget corresponding to $c$ is positioned to the left of the gadget of $c'$.
We add an interval whose left endpoint is in the interval corresponding to $x_c''$ and whose right endpoint is in the interval of $x_{c'}'$.
The vertices corresponding to these intervals will be called \emph{connector vertices}.
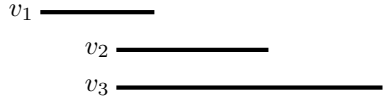
\begin{figure}[t]
    \centering

    \begin{tikzpicture}
        \gadgetLeftSmall{0}{0}{c};
    \end{tikzpicture}

    \caption{Left gadget.}
    \label{fig:leftGadget}
\end{figure}
Also, add the \emph{left gadget} illustrated in \Cref{fig:leftGadget} to the very left, not intersecting any of the intervals of the clause gadgets nor any of the connector vertices.
Now, for each clause $c\in C$ we add two more intervals as follows.
Add an interval whose left endpoint lies in the interval of $v_2$ but not in the interval of $v_1$, and whose right endpoint lies in the interval corresponding to $v_c$ but to the left of any variable vertex in the clause gadget of $c$.
The corresponding vertex of this interval is called the \emph{opening vertex of $c$}.
Add another interval whose left endpoint lies in the interval corresponding to $v_3$ but not in the interval of $v_2$, and whose right endpoint lies in the interval corresponding to $v_c$ but to the right of every variable vertex in the clause gadget of $c$.
We call the corresponding vertex the \emph{closing vertex of $c$}.
In \Cref{fig:example}, we show a simple example with three clauses in which only one variable appears twice (the right-most positioned variable in the middle clause and the middle variable in the right-most clause).

Notice that $G_{I}$ can be constructed in polynomial time from a given instance $I$.
We claim that $G_{I}$ has a proper coloring if and only if $I$ is satisfiable.

First, assume we have a proper coloring $col$ of $G_{I}$.
It is easy to see that $v_1$ is the only vertex colored with 1 (because it is the only vertex of degree 2) and the vertices $v_2$ and $v_3$ are the only ones colored with 2 (because they are the only neighbors of a vertex of color 1).
Thus, all the opening vertices are colored with 3, and all the closing vertices are colored with 4, and no other vertices can get these colors.
Now notice that every clause vertex is adjacent to the same number of opening vertices and closing vertices, but this property does not hold for any variable, value or connector vertex (the connector vertices are adjacent to the same opening and closing vertices as the value vertex in which its left endpoint is contained).
Thus, all clause vertices are colored with 5 and no other vertices are.
Since each connector vertex has at least two clause vertices in its neighborhood, these vertices receive color 6.
All the variable and value vertices have exactly one neighbor of color 5, therefore, they are colored with 7, \texttt{T}, \texttt{t}, \texttt{F} or \texttt{f}.
Then we have that each of the value vertices has exactly one neighbor which is not colored with a color from 1 to 6, while the variable vertices have two of them.
Thus, variable vertices are colored with 7, and value vertices are colored with \texttt{T}, \texttt{t}, \texttt{F} or \texttt{f}.
For every variable vertex, the set of colors assigned to their two neighboring value vertices is either $\set{\texttt{T}, \texttt{t}}$ or $\set{\texttt{F}, \texttt{f}}$.
Now consider the value vertices in the neighborhood of some fixed connector vertex $w$.
We have shown that all of them, except for those whose intervals contain the endpoints of $w$, appear in pairs colored either with \texttt{T} and \texttt{t}, or with \texttt{F} and \texttt{f}.
Since the number of value vertices in $N(w)$ which receive color \texttt{T} must be the same as the number of value vertices in $N(w)$ with color \texttt{t} (and analogously for \texttt{F} and \texttt{f}), it follows that the two value vertices containing the endpoints of $w$ must either receive the colors \texttt{T} and \texttt{t} or the colors \texttt{F} and \texttt{f}.
This implies that for every variable $x\in X$ either for every clause $c$ which contains $x$ we have that $col(x'_c),col(x''_c)\in\set{\texttt{T},\texttt{t}}$, or for every clause $c$ containing $x$ we have that $col(x'_c),col(x''_c)\in\set{\texttt{F},\texttt{f}}$.
Finally, every vertex of color 5 has exactly one neighbor of color \texttt{T}.
This means that in the neighborhood of every clause vertex there is exactly one variable vertex whose adjacent value vertices are colored with $\set{\texttt{T}, \texttt{t}}$, and the remaining two variables vertices have their adjacent value vertices colored with $\set{\texttt{F}, \texttt{f}}$.
We set a variable $x\in X$ to \True\ if all value vertices adjacent to the variable vertices corresponding to a literal of $x$ are colored with \texttt{T} or \texttt{t} and to \False\ otherwise.
We claim that this yields a satisfying assignment of $I$.
Indeed, each variable gets a single value (either $\True$ or $\False$, since the set of colors of corresponding value vertices is either $\set{\texttt{T}, \texttt{t}}$ or $\set{\texttt{F}, \texttt{f}}$), and each clause has exactly a variable set to $\True$.

Assume now that the given instance $I$ of positive 1-in-3 SAT is satisfiable.
Let $f$ be a function from the set of variables to $\set{\True, \False}$ that assigns values to the variables satisfying the instance.
We will construct a proper coloring of the graph $G_I$.
Assign color 1 to $v_1$, color 2 to $v_2$ and $v_3$, color 3 to every opening vertex, color 4 to every closing vertex, color 5 to every clause vertex, color 6 to every connector vertex, and color 7 to every variable vertex.
For every value vertex $x_c'$, if $f(x) = \True$ then assign $x_c'$ the color \texttt{T}, otherwise assign $x_c'$ the color \texttt{F}.
Analogously, for every value vertex $x_c''$, if $f(x) = \True$ assign \highlightChange{$x_c''$} color \texttt{t}, otherwise assign \highlightChange{$x_c''$} color \texttt{f}.
It is easy to verify that this is a proper coloring.

\section{Applications}
\label{sec:Applications}
In this section, we give some examples of problems that can be modeled as $d$-stable locally checkable problems with a bounded number of colors, some of them with size and connectivity constraints on the color classes.
Their complexity was previously unknown on graphs of bounded mim-width and, as a consequence of the results in \Cref{sec:Algorithm} (as well as in \Cref{sec:LCPtoDN}), all these problems are XP parameterized by
\highlightChange{the mim-width of a given binary decomposition tree of the input graph}.

Independently,
\highlightChange{Bergougnoux, Dreier and Jaffke~\cite{DN-logic}}
give a DN logic formula for conflict-free coloring and $b$-coloring with a bounded number of colors.
Their expression is shorter and more elegant than the one that can be obtained from \Cref{sec:LCPtoDN}.
However, applying the algorithm of \cite{DN-logic} to these formulas yields a slower runtime than the algorithm in \Cref{sec:Algorithm}.
Although every problem that is expressible as a $d$-stable locally checkable problem can be expressed in DN-logic too, we believe that for some of these problems it is more natural to express them as a $d$-stable locally checkable problem.

\subsection{\texorpdfstring{$[k]$}{[k]}-Roman domination}

This problem was first defined in~\cite{tripleRoman} as a generalization of Roman domination and double Roman domination~\cite{RomanDomination,doubleRoman}. 
Let $k$ be a positive integer.
A \emph{$[k]$-Roman dominating function} on a graph $G$ is a function $f \colon V(G) \to \set{0, \ldots, k+1}$
which has the property that if $f(v) < k$ then
$\sum_{u \in N_G[v]} f(u) \geq |AN_G^f(v)| + k$,
where $AN_G^f(v) = \{u \in N_G(v) : f(u) \geq 1\}$ (this set is called the \emph{active neighborhood of $v$}).
The \emph{weight} of a $[k]$-Roman dominating function $f$ is $\sum_{v \in V(G)} f(v)$, and the minimum weight of a $[k]$-Roman dominating function on $G$ is the \emph{$[k]$-Roman domination number} of $G$.
The \textsc{$[k]-$Roman domination} problem consists in computing the $[k]$-Roman domination number of a given graph.

In~\cite{LCPcliquewidth}, \textsc{$[k]-$Roman domination} was shown to be solvable in linear time on graphs of bounded clique-width by modeling it as a locally checkable problem with a color-counting check function, as follows:
\begin{itemize}
	\item $\ColorSet = \set{0, \ldots, k+1}$,

	\item $(\WeightSet, \wlesseq, \weightsSum) = (\mathbb{N} \cup \{+\infty\}, \leq, +)$,

	\item $\textsc{w}(v, a, \ell_0, \ldots, \ell_{k+1}) = a$ for all $v \in V(G)$, $a \in \ColorSet$ and $\ell_0, \ldots, \ell_{k+1} \in \N$,

	\item $check(v, a, \ell_0, \ldots, \ell_{k+1}) = \left(a + \sum_{j=0}^{k+1} j\ell_j \geq k + \sum_{j=1}^{k+1} \ell_j\right)$ for all $v \in V(G)$, $a \in \ColorSet$ and $\ell_0, \ldots, \ell_{k+1} \in \N$.
\end{itemize}

Notice that $\textsc{w}$ is 1-stable and $check$ is $k$-stable, so this model also suffices our purposes.
Furthermore, as mentioned in~\cite{LCPTreewidth}, by making slight changes in the model (that do not affect the $k$-stability of the functions or the number of colors by more than a constant factor) we can also express several other variants of Roman domination, such as perfect~\cite{perfectDoubleRoman}, independent~\cite{independentDoubleRoman}, outer independent~\cite{outerindepDoubleRoman}, total~\cite{totalDoubleRoman} and maximal~\cite{LCPTreewidth}.
For the signed~\cite{signedRoman} and
signed total~\cite{signedTotalRoman} Roman domination, notice that they are color-counting, therefore FPT parameterized by clique-width~\cite{LCPcliquewidth}, however they are not $d$-stable for any~$d$.

\subsection{Dual domination}

\def\inD{\textsc{d}}
\def\inNDVplus{\textsc{n}^{+}}
\def\inNDVminus{\textsc{n}^{-}}
\def\theRest{\overline{\textsc{n}}}

\highlightChange{In~\cite{dualDomination}, three recent variations of the domination problem were defined.}
Given a graph partitioned into the \emph{positive} vertices ($V^{+}$) and the \emph{negative} vertices ($V^{-}$), in each problem we want to find a subset of positive vertices $D \subseteq V^{+}$ with conditions as follow:
\begin{itemize}
\item \textsc{Maximum Bounded Dual Domination} (MBDD):
for an integer $k$ given as input, $|N(D) \cap V^{-}| \leq k$ and $|N[D] \cap V^{+}|$ is as large as possible.

\item \textsc{Maximum Dual Domination} (MDD):
$D$ maximizes $|N[D] \cap V^{+}| - |N(D) \cap V^{-}|$.

\item \textsc{Minimum Negative Dual Domination} (mNDD):
$D$ dominates $V^{+}$ (that is, $V^{+} \subseteq N[D]$) and minimizes $|N(D) \cap V^{-}|$.
\end{itemize}
All three problems are NP-hard and there exist linear-time algorithms to solve MDD and mNDD on trees, as well as a $O(k^2 |V(G)|)$-time algorithm to solve MBDD on trees~\cite{dualDomination}.

We will show that all three problems can be modeled as $1$-stable locally checkable problems.
In these models, the set $D$ which we are searching for is the color class $\inD$.

\textsc{Maximum Bounded Dual Domination} (MBDD):
\begin{itemize}
	\item $\ColorSet = \set{\inD, \inNDVplus, \inNDVminus, \theRest}$,

	\item $(\WeightSet, \wlesseq, \weightsSum) = (\mathbb{N} \cup \{-\infty\}, \geq, +)$,

	\item for all $v \in V(G)$, $a \in \ColorSet$ and $\ell_{\inD}, \ell_{\inNDVplus}, \ell_{\inNDVminus}, \ell_{\theRest} \in \N$ we have:
    \[
        \textsc{w}(v, a, \ell_{\inD}, \ell_{\inNDVplus}, \ell_{\inNDVminus}, \ell_{\theRest}) =
            \begin{cases}
            1   & \text{if } a \in \set{\inD, \inNDVplus} \\
            0   & \text{otherwise,}
            \end{cases}
    \]

	\item for all $v \in V(G)$, $a \in \ColorSet$ and $\ell_{\inD}, \ell_{\inNDVplus}, \ell_{\inNDVminus}, \ell_{\theRest} \in \N$ we have:
    \begin{itemize}
        \item $check(v, \inD, \ell_{\inD}, \ell_{\inNDVplus}, \ell_{\inNDVminus}, \ell_{\theRest}) = (v \in V^{+})$
        \item $check(v, \inNDVplus, \ell_{\inD}, \ell_{\inNDVplus}, \ell_{\inNDVminus}, \ell_{\theRest}) = (v \in V^{+} \land \ell_{\inD} \geq 1)$
        \item $check(v, \inNDVminus, \ell_{\inD}, \ell_{\inNDVplus}, \ell_{\inNDVminus}, \ell_{\theRest}) = (v \in V^{-} \land \ell_{\inD} \geq 1)$
        \item $check(v, \theRest, \ell_{\inD}, \ell_{\inNDVplus}, \ell_{\inNDVminus}, \ell_{\theRest}) = (\ell_{\inD} = 0)$,
    \end{itemize}

    \item Size constraint: the number of vertices of color $\inNDVminus$ is at most $k$.
\end{itemize}

\textsc{Maximum Dual Domination} (MDD):
\begin{itemize}
	\item $\ColorSet = \set{\inD, \inNDVplus, \inNDVminus, \theRest}$,

	\item $(\WeightSet, \wlesseq, \weightsSum) = (\mathbb{Z} \cup \{-\infty\}, \geq, +)$,

	\item for all $v \in V(G)$, $a \in \ColorSet$ and $\ell_{\inD}, \ell_{\inNDVplus}, \ell_{\inNDVminus}, \ell_{\theRest} \in \N$ we have:
    \[
        \textsc{w}(v, a, \ell_{\inD}, \ell_{\inNDVplus}, \ell_{\inNDVminus}, \ell_{\theRest}) =
            \begin{cases}
            1   & \text{if } a \in \set{\inD, \inNDVplus} \\
            -1  & \text{if } a = \inNDVminus \\
            0   & \text{otherwise,}
            \end{cases}
    \]

	\item for all $v \in V(G)$, $a \in \ColorSet$ and $\ell_{\inD}, \ell_{\inNDVplus}, \ell_{\inNDVminus}, \ell_{\theRest} \in \N$ we have:
    \begin{itemize}
        \item $check(v, \inD, \ell_{\inD}, \ell_{\inNDVplus}, \ell_{\inNDVminus}, \ell_{\theRest}) = (v \in V^{+})$
        \item $check(v, \inNDVplus, \ell_{\inD}, \ell_{\inNDVplus}, \ell_{\inNDVminus}, \ell_{\theRest}) = (v \in V^{+} \land \ell_{\inD} \geq 1)$
        \item $check(v, \inNDVminus, \ell_{\inD}, \ell_{\inNDVplus}, \ell_{\inNDVminus}, \ell_{\theRest}) = (v \in V^{-} \land \ell_{\inD} \geq 1)$
        \item $check(v, \theRest, \ell_{\inD}, \ell_{\inNDVplus}, \ell_{\inNDVminus}, \ell_{\theRest}) = (\ell_{\inD} = 0)$.
    \end{itemize}
\end{itemize}

\textsc{Minimum Negative Dual Domination} (mNDD):
\begin{itemize}
	\item $\ColorSet = \set{\inD, \inNDVplus, \inNDVminus, \theRest}$,

	\item $(\WeightSet, \wlesseq, \weightsSum) = (\mathbb{N} \cup \{+\infty\}, \leq, +)$,

	\item for all $v \in V(G)$, $a \in \ColorSet$ and $\ell_{\inD}, \ell_{\inNDVplus}, \ell_{\inNDVminus}, \ell_{\theRest} \in \N$ we have:
    \[
        \textsc{w}(v, a, \ell_{\inD}, \ell_{\inNDVplus}, \ell_{\inNDVminus}, \ell_{\theRest}) =
            \begin{cases}
            1  & \text{if } a = \inNDVminus \\
            0   & \text{otherwise,}
            \end{cases}
    \]

	\item for all $v \in V(G)$, $a \in \ColorSet$ and $\ell_{\inD}, \ell_{\inNDVplus}, \ell_{\inNDVminus}, \ell_{\theRest} \in \N$ we have:
    \begin{itemize}
        \item $check(v, \inD, \ell_{\inD}, \ell_{\inNDVplus}, \ell_{\inNDVminus}, \ell_{\theRest}) = (v \in V^{+})$
        \item $check(v, \inNDVplus, \ell_{\inD}, \ell_{\inNDVplus}, \ell_{\inNDVminus}, \ell_{\theRest}) = (v \in V^{+} \land \ell_{\inD} \geq 1)$
        \item $check(v, \inNDVminus, \ell_{\inD}, \ell_{\inNDVplus}, \ell_{\inNDVminus}, \ell_{\theRest}) = (v \in V^{-} \land \ell_{\inD} \geq 1)$
        \item $check(v, \theRest, \ell_{\inD}, \ell_{\inNDVplus}, \ell_{\inNDVminus}, \ell_{\theRest}) = (v \in V^{-} \land \ell_{\inD} = 0)$.
    \end{itemize}
\end{itemize}

The models above imply that these three problems are XP when parameterized by \highlightChange{the mim-width of a given binary decomposition tree of the input graph}.
Moreover, by the results in~\cite{LCPcliquewidth}, they are also FPT when parameterized by clique-width.

\subsection{Happy colorings}

In a vertex coloring $c$ of a graph $G$, a vertex $v$ is called \emph{happy} if all its neighbors receive the color $c(v)$.
Given a partial coloring of a graph \highlightChange{$G$} using $k$ colors \highlightChange{(that is, a coloring $p$ of domain $P$ for some $P \subseteq V(G)$)}, we want to determine the maximum number of happy vertices we can obtain in an extension of this coloring \highlightChange{(that is, in a coloring $c$ of the whole graph $G$, where $c$ restricted to $P$ is the coloring $p$)}.
This problem is known as \textsc{$k$-Maximum Happy Vertices} ($k$-MHV) and was first defined in~\cite{happyColoringDef}.
In the same paper it is proven that, for every $k \geq 3$, this problem is NP-complete.
Algorithmic results for this problem so far include: an FPT algorithm parameterized by treewidth and neighborhood diversity~\cite{happyColoringTreewidth}, and an FPT algorithm parameterized by clique-width~\cite{happyColoringCliquewidth}.
Also, a natural generalization to weighted graphs is defined in~\cite{happyColoringTreewidth} (that is, there is a function $w \colon V(G) \to \N$ and we want to maximize the sum of $w(v)$ over all the happy vertices $v$).
We can model \highlightChange{weighted} $k$-MHV as a 1-stable locally checkable problem, as follows:
\begin{itemize}
	\item $\ColorSet = \set{1, \ldots, k}$,

	\item for all $v \in V(G)$, $a \in \ColorSet$ and $\ell_1, \ldots, \ell_{k} \in \N$:
    \[
        check(v, a, \ell_1, \ldots, \ell_{k}) =
            \begin{cases}
                \False  & \text{if $v \in P$ and $a \neq p(v)$} \\
                \True       & \text{otherwise\highlightChange{,}}
            \end{cases}
    \]
    where $P$ is the set of precolored vertices and $p(v)$ is the color received by $v$ in the precoloring,

	\item $(\WeightSet, \wlesseq, \weightsSum) = (\mathbb{N} \cup \{-\infty\}, \geq, +)$,

	\item for all $v \in V(G)$, $a \in \ColorSet$ and $\ell_1, \ldots, \ell_{k} \in \N$ we have:
    $$\textsc{w}(v, a, \ell_1, \ldots, \ell_{k}) =
        \begin{cases}
        w(v)    &   \text{if $\ell_i = 0$ for all $i \in [k] - \{a\}$}\\
        0       &   \text{otherwise.}
        \end{cases}$$
    Notice that this function is $1$-stable.
\end{itemize}

The \textsc{Maximum Happy Set} problem (MaxHS), defined in~\cite{happySetProblem}, asks for a subset $S$ of $k$ vertices that maximizes the number of vertices $v \in S$ such that $N[v] \subseteq S$.
This problem is NP-hard, as well as FPT when parameterized by either tree-width, by neighborhood diversity, or by cluster deletion number~\cite{happySetProblem}.
It is also FPT when parameterized by modular-width or by clique-width~\cite{happySetCliquewidth}, and polynomial-time solvable on interval graphs~\cite{happySetInterval}.
With the following model we show that MaxHS is $1$-stable with 2 colors and a size constraint.
\begin{itemize}
	\item $\ColorSet = \set{\textsc{s}, \overline{\textsc{s}}}$,

	\item for all $v \in V(G)$, $a \in \ColorSet$ and $\ell_{\textsc{s}}, \ell_{\overline{\textsc{s}}} \in \N$:
    $check(v, a, \ell_{\textsc{s}}, \ell_{\overline{\textsc{s}}}) = \True$,

	\item $(\WeightSet, \wlesseq, \weightsSum) = (\mathbb{N} \cup \{-\infty\}, \geq, +)$,

	\item for all $v \in V(G)$, $a \in \ColorSet$ and $\ell_1, \ldots, \ell_{k} \in \N$ we have:
    $$\textsc{w}(v, a, \ell_{\textsc{s}}, \ell_{\overline{\textsc{s}}}) =
        \begin{cases}
        1   &   \text{if $a = \textsc{s}$ and $\ell_{\overline{\textsc{s}}} = 0$}\\
        0   &   \text{otherwise.}
        \end{cases}$$

    \item The size of the color class $\textsc{s}$ is $k$.
\end{itemize}

\subsection{Locally bounded coloring}

Suppose we are given a graph $G$, two positive integers $p$ and $k$, a partition of the vertex set $V(G)$ into $p$ subsets $V_1, \ldots, V_p$, and $pk$ integral bounds $n_{i,j}$, with $i \in [p]$ and $j \in [k]$, such that $\sum_{j=1}^{k} n_{i,j} = |V_i|$ for each $i \in [p]$.
The \textsc{Locally Bounded $k$-Coloring} problem, defined in~\cite{LocallyBoundedColoring}, consists in deciding if there exists a coloring of $G$ using $k$ colors such that no two \highlightChange{adjacent vertices} get the same color and that, for each $i \in [p]$ and for each $j \in [k]$, the number of vertices having color $j$ in $V_i$ is $n_{i,j}$.
For fixed values of $p$ and $k$, this problem is NP-complete~\cite{LocallyBoundedColoring} and polynomial-time solvable on graphs of bounded tree-width and on cographs~\cite{LocallyBoundedColoring2}.

We propose the following model for Locally Bounded $k$-Coloring, with fixed $p$ and $k$, as a $1$-stable locally checkable problem using $pk$ colors and using size constraints.
\begin{itemize}
	\item $\ColorSet = [p] \times [k]$,

	\item $(\WeightSet, \wlesseq, \weightsSum) = (\set{0,1}, \leq, \max)$,

	\item $\textsc{w}(v, (i,j), \ell_{(1,1)}, \ldots, \ell_{(p,k)}) = 0$
    for all $v \in V(G)$, $(i,j) \in \ColorSet$ and all $\ell_{(i',j')} \in \N$ with $(i',j') \in \ColorSet$,

	\item $check(v, (i,j), \ell_{(1,1)}, \ldots, \ell_{(p,k)}) = \left(
        v \in V_i
        \land
        \sum_{i' \in [p]} \ell_{(i',j)} = 0
    \right)$
    for all $v \in V(G)$, $a \in \ColorSet$ and all $\ell_{(i',j')} \in \N$ with $(i',j') \in \ColorSet$,

    \item Size constraints: for each pair $(i,j) \in \ColorSet$, the number of vertices of color $(i,j)$ is $n_{i,j}$.
\end{itemize}

\subsection{Conflict-free \texorpdfstring{$k$}{k}-coloring}

Conflict-free coloring problems are a group of problems in all of which we look for a coloring of a graph such that in the neighborhood of every vertex there is a color which appears exactly once.
The first such problem was defined in~\cite{CF-def}.
The following variations have been studied as well (for example in~\cite{cf-free-clique}).
A \emph{CFON $k$-coloring} of a graph $G$ is a coloring $c \colon V(G) \rightarrow [k]$ such that for every vertex $v \in V(G)$ there is at least one vertex $u \in N(v)$ such that no other vertex in $N(v)$ has color $c(u)$.
A \emph{CFCN $k$-coloring} of a graph $G$ is a coloring $c \colon V(G) \rightarrow [k]$ such that for every vertex $v \in V(G)$ there is at least one vertex $u \in N[v]$ such that no other vertex in $N[v]$ has color $c(u)$.
A \emph{CFON$^*$ $k$-coloring} of a graph $G$ is a coloring $c \colon V(G) \rightarrow \set{0, \ldots, k}$ such that for every vertex $v \in V(G)$ there exists at least one vertex $u \in N(v)$ with $c(u) \neq 0$ and such that no other vertex in $N(v)$ has color $c(u)$.
A \emph{CFCN$^*$ $k$-coloring} of a graph $G$ is a coloring $c \colon V(G) \rightarrow \set{0, \ldots, k}$ such that for every vertex $v \in V(G)$ there exists at least one vertex $u \in N[v]$ with $c(u) \neq 0$ and such that no other vertex in $N[v]$ has color $c(u)$.
For each of these variants we want to determine the smallest $k$ for which such a $k$-coloring exists.
Each of these problems is NP-complete on general graphs.

We can model the CFON$^*$ $k$-coloring problem as a 2-stable locally checkable problem with $k+1$ colors:
\begin{itemize}
	\item $\ColorSet = \set{0, \ldots, k}$,

	\item $(\WeightSet, \wlesseq, \weightsSum) = (\set{0,1}, \leq, \max)$,

	\item $\textsc{w}(v, a, \ell_0, \ldots, \ell_k) = 0$ for all $v \in V(G)$, $a \in \ColorSet$ and $\ell_0, \ldots, \ell_k \in \N$,

    \item $check(v, a, \ell_0, \ldots, \ell_k) =( \existsFormula{j \in [k] }{\ell_j=1})$ for all $v \in V(G)$, $a \in \ColorSet$ and $\ell_0, \ldots, \ell_k \in \N$.
\end{itemize}

For all the other variants of this problem we can easily modify the check function to model them correctly while preserving the 2-stability.

It was shown in~\cite{cf-free-clique} that the problem of finding a CFCN$^*$- and a CFON$^*$-coloring using the minimum number of colors is fixed-parameter tractable with respect to the combined parameters clique-width and number of colors $k$.
Our result shows that the problems are solvable in polynomial time when mim-width and number of colors are bounded.
In \cite{cf-free-clique} they further show that any interval graph has a CFON$^*$ 3-coloring and ask for \highlightChange{a polynomial-time} algorithm determining the smallest number of colors needed for a CFON$^*$ coloring of an interval graph.
Our result gives such an algorithm.

\subsection{b-coloring with fixed number of colors}

Given a graph $G$, a \highlightChange{\emph{b-coloring}} of $G$ is a coloring of $V(G)$ such that no pair of neighbors receive the same color and also every color class contains a vertex that has neighbors in all the other color classes.
In \cite{b-chromatic-number}, Irving and Manlove define the b-chromatic number of $G$, as the maximum $k$ such that $G$ admits a b-coloring with $k$ colors, and prove that the problem of determining such number is NP-complete for general graphs.
The b-coloring problem asks, for a given graph $G$ and an integer $k$, if $G$ admits a b-coloring with $k$ colors.
It is shown in~\cite{b-coloring-NPc} that this problem is NP-complete even for bipartite graphs.
However, it has been shown in~\cite{b-coloring-cw} that the b-coloring problem can be solved in polynomial time on graphs of bounded clique-width, and in~\cite{b-coloring-thinness} that for a fixed number of colors it can be solved in polynomial time on graphs of bounded proper thinness.
Here we show that the b-coloring problem with a fixed number of colors $k$ can be modeled with the following 1-stable locally checkable problem with $2k$ colors:
\begin{itemize}
	\item $\ColorSet = \set{-k, \ldots, -1, 1, \ldots, k}$,

	\item $(\WeightSet, \wlesseq, \weightsSum) = (\set{0,1}, \leq, \max)$,

	\item $\textsc{w}(v, a, \ell_{-k}, \ldots, \ell_k) = 0$
	for all $v \in V(G)$, $a \in \ColorSet$ and all $\ell_b \in \N$ with $b \in \ColorSet$,

	\item for every vertex $v \in V(G)$, every $a \in [k]$ and every $\ell_b \in \N$ with $b \in \ColorSet$, we have that
	$check(v, a, \ell_{-k}, \ldots, \ell_k) = (\ell_{a} + \ell_{-a} = 0)$
	and
	$check(v, -a, \ell_{-k}, \ldots, \ell_k) = (\ell_{a} + \ell_{-a} = 0 \land \forallFormula{i \in [k] \setminus \set{a}}{\ell_i + \ell_{-i}\geq 1})$,

	\item For each $a \in [k]$, the size of the color class $-a$ is at least 1.
\end{itemize}

It is easy to see that that a proper coloring satisfying the size constraint corresponds to a b-coloring, where the vertices colored with negative integers represent the ``b-vertices'' (that is, the vertices that have at least one neighbor of each other color).

\section*{Acknowledgements}
\label{sec:ack}
Carolina L. Gonzalez is partially supported by a CONICET doctoral fellowship, CONICET PIP 11220200100084CO,
UBACyT 20020170100495BA and ANPCyT PICT-2021-I-A-00755.

\highlightChange{We would like to thank the anonymous reviewers for their valuable comments that helped improve our manuscript.}

\bibliographystyle{elsarticle-harv}
\biboptions{sort&compress}
\bibliography{Z-References}

\end{document}